\documentclass[10pt,twocolumn,aps,pra,showpacs,superscriptaddress,floatfix,longbibliography,nofootinbib]{revtex4-1}

\usepackage[T1]{fontenc}
\usepackage{graphicx}
\usepackage[utf8]{inputenc}
\usepackage[T1]{fontenc}
\usepackage[sc,osf]{mathpazo}
\usepackage{amsmath, amsthm, amssymb,amsfonts}
\usepackage{color}
\usepackage{psfrag}
\usepackage{epsfig}
\usepackage{bbm}
\usepackage{bm}
\usepackage[colorlinks=true,citecolor=blue,urlcolor=magenta]{hyperref}
\usepackage[normalem]{ulem}
\usepackage{epstopdf}
\usepackage{graphicx}
\usepackage{stackengine,xcolor}

\newcommand{\red}[1]{{\color{red} #1}}

\definecolor{nred}{rgb}{0.9,0.1,0.1}
\definecolor{nblack}{rgb}{0,0,0}
\definecolor{nblue}{rgb}{0.2,0.2,0.8}
\definecolor{ngreen}{rgb}{0.2,0.6,0.2}

\newcommand{\blu}{\color{nblue}}
\newcommand{\grn}{\color{ngreen}}

\usepackage{etoolbox}
\usepackage{bbold} 

\newcommand{\ket}[1]{| #1 \rangle}
\newcommand{\bra}[1]{\langle #1 |}

\newcommand{\IR}{\mathcal{IR}}
\newcommand{\IRJ}{\mathcal{IR}^{{\tiny \rm J}}}
\newcommand{\IRR}{\mathcal{IR}^{{\tiny \rm R}}}
\newcommand{\IRP}{\mathcal{IR}^{{\tiny \rm P}}}
\newcommand{\DIIR}{\mathcal{IR_{\rm DI}}}

\newcommand{\IW}{\mathcal{IW}}

\newcommand{\beq}{\begin{eqnarray}}
\newcommand{\eeq}{\end{eqnarray}}

\newcommand{\mean}[1]{\langle #1 \rangle}

\newcommand{\rab}{{\varrho^{\text{AB}}}}

\newcommand{\Eax}{{E^\text{A}_{a|x}}}
\newcommand{\Eby}{{E^\text{B}_{b|y}}}

\DeclareMathOperator{\tr}{tr}

\theoremstyle{definition}

\AtBeginDocument{%
    \newwrite\bibnotes
    \def\bibnotesext{Notes.bib}
    \immediate\openout\bibnotes=\jobname\bibnotesext
    \immediate\write\bibnotes{@CONTROL{REVTEX41Control}}
    \immediate\write\bibnotes{@CONTROL{%
    apsrev41Control,author="08",editor="1",pages="1",title="0",year="1"}}
     \if@filesw
     \immediate\write\@auxout{\string\citation{apsrev41Control}}%
    \fi
}%

\begin{document}

\title{Device-independent quantification of measurement incompatibility}

\author{Shin-Liang Chen}
\email{shin.liang.chen@phys.ncku.edu.tw}
\affiliation{Dahlem Center for Complex Quantum Systems, Freie Universit\"at Berlin, 14195 Berlin, Germany}
\affiliation{Department of Physics, National Cheng Kung University, Tainan 701, Taiwan}
\affiliation{Center for Quantum Frontiers of Research \& Technology (QFort), National Cheng Kung University, Tainan 701, Taiwan}

\author{Nikolai Miklin} 
\email{nikolai.miklin@ug.edu.pl}
\affiliation{International Centre for Theory of Quantum Technologies (ICTQT), University of Gdansk, 80-308 Gda\'nsk, Poland}

\author{Costantino Budroni}
\email{costantino.budroni@univie.ac.at}
\affiliation{Faculty of Physics, University of Vienna, Boltzmanngasse 5, 1090 Vienna, Austria}
\affiliation{Institute for Quantum Optics and Quantum Information (IQOQI), Austrian Academy of Sciences, Boltzmanngasse 3, 1090 Vienna, Austria}

\author{Yueh-Nan Chen}
\email{yuehnan@mail.ncku.edu.tw }
\affiliation{Department of Physics, National Cheng Kung University, Tainan 701, Taiwan}
\affiliation{Center for Quantum Frontiers of Research \& Technology (QFort), National Cheng Kung University, Tainan 701, Taiwan}

\date{ \today}

\begin{abstract}
Incompatible measurements, i.e., measurements that cannot be simultaneously performed, are necessary to observe nonlocal correlations. It is natural to ask, e.g., how incompatible the measurements have to be to achieve a certain violation of a Bell inequality. In this work, we provide the direct link between Bell nonlocality and the quantification of measurement incompatibility.  This includes quantifiers for both incompatible and genuine-multipartite incompatible measurements. Our method straightforwardly generalizes to include constraints on the system's dimension (semi-device-independent approach) and on projective measurements, providing improved bounds on incompatibility quantifiers, and to include the prepare-and-measure scenario.
\end{abstract}

\maketitle

\section{Introduction}
One of the most intriguing phenomena in quantum theory is that there exist physical quantities whose values cannot be simultaneously obtained. The most celebrated example is arguably the position and momentum of a particle, initially formulated in terms of the uncertainty relation~\cite[]{Heisenberg1927,Robertson29,BLWRev}. 
Such a phenomenon, called \emph{measurement incompatibility} (or simply \emph{incompatibility}), enables one to demonstrate several remarkable quantum features such as quantum nonlocality~\cite[]{Bell64,Brunner14}, quantum steering~\cite[]{Schrodinger35,Cavalcanti17,Uola2020Steering}, quantum contextuality~\cite[]{KS67,KlyachkoPRL2008,CabelloPRL2008,LIANG2011,Context_review} (see, respectively, \cite[]{Wolf09}, \cite[]{Quint14,Uola14}, and \cite[]{Xu19,Tavakoli2020Measurement}), and provides a resource to many quantum information protocols (see, e.g.,~\cite[]{Carmeli19,Skrzypczyk19,Uola19a,Takagi19PRL,Takagi19PRX,Oszmaniec2019operational,Mori2019Operational,Buscemi2020Complete}).
In a more modern language, incompatibility has been formulated as the non-existence of a \emph{joint measurement}~\cite[]{Lahti03}. 

Nonlocality plays a central role in quantum information (QI), more precisely, in the definition of device-independent (DI) QI~\cite[]{Acin07,Scarani12,Brunner14}: without any characterization of the measurement devices (e.g., measurement operators, states, system dimension) all information is encoded in $P(a,b|x,y)$, the probability of the outputs $a,b$ given the measurement settings $x,y$. Provided that $P(a,b|x,y)$ is nonlocal, a surprisingly high variety of statements and QI protocols can be based on such correlations: from quantum key distribution~\cite{Acin07}, to entanglement detection~\cite[]{Moroder13}, randomness certification~\cite[]{Pironio10}, verification of steerability~\cite[]{Wiseman07,Cavalcanti16,CBLC16,CBLC18}, witnessing dimension of quantum systems~\cite[]{Gallego10}, and so on.

In this sense, a violation of a Bell inequality is also a DI witness of incompatibility, as incompatible measurements are necessary to observe it~\cite[]{Wolf09,HirschPRA2018,BeneNJP2018}. In this work, we address the quantitative question: How incompatible the underlying measurements have to be in order to observe a certain quantum violation of a Bell inequality? A central tool in our investigation is the notion of {\it moment matrix}, that has wide applications in the characterization of quantum correlations and DI approaches ~\cite{Doherty08,NPA,Pironio10b,Moroder13}. 
Here, we introduce the {\it measurement moment matrix} (MMM) that allows us to quantify several quantities that are formulated via semidefinite programming (SDP)~\cite{BoydBook} in terms of measurement effects, such as \emph{incompatibility robustness}~\cite[]{Haapasalo15Robustness,Uola15,Heinosaari16}, \emph{genuine-multipartite incompatibility}~\cite[]{QuintinoPRL2019},  and similar quantities \cite[]{Pusey15,Heinosaari15,Cavalcanti16}. 

\begin{figure}[t]
\emph{\includegraphics[width=0.45\textwidth]{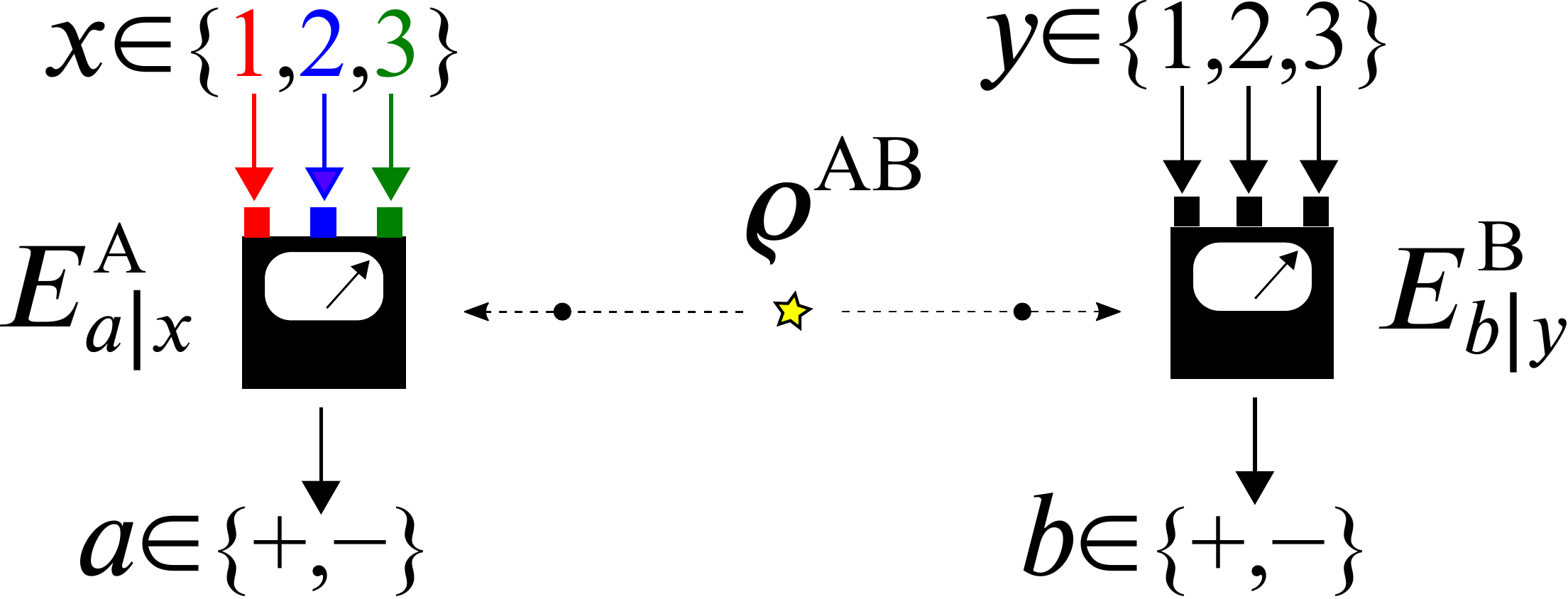} }
\caption{Schematic representation of device-independent quantification of measurement incompatibility. By performing measurements on two distant particles and observing the correlations $P(a,b|x,y)$, we are able to estimate several measures of incompatiblity among pairs, $(\red{1},{\blu 2})$, $(\red{1},{\grn 3})$, $(\blu{2},{\grn 3})$, or triples, $(\red{1},{\blu 2},{ \grn 3})$, or genuine triplewise incompatibility among  $(\red{1}, {\blu 2}, {\grn 3})$.}
\label{Fig_picture}
\end{figure}

Our results allows for investigations beyond the DI scenario. In fact, due to its generality the idea of MMMs can be straightforwardly extended to the semi-DI approach, i.e., where the dimension of quantum system is assumed to be known~\cite[]{PawlowskiPRA2011,LiangPRA2011}, to investigate the role of dimension constraints or even non-projectivness in measurement incompatibility, and it can be extended even to the prepare-and-measure scenario.

\section{Incompatible measurements}
Let us start by briefly reviewing the concept of measurement incompatibility. Consider a quantum measurement described by a {\it positive-operator-valued measure} (POVM) $\{\Eax\}_a$ for a given $x$, where the indices $x\in\mathcal{X}$ and $a\in\mathcal{A}$, label the measurement settings and outcomes of the measurement,  respectively. The operators $\Eax$, called {\it effect operators} are positive semidefinite, i.e., $\Eax\succeq 0~\forall a,x$, and satisfy the normalization condition, $\sum_a \Eax=\openone~\forall x$. A collection of POVMs $\{\Eax\}_{a,x}$ is called a \emph{measurement assemblage}~\cite[]{Piani15}. A measurement assemblage is said to be \emph{compatible} or \emph{jointly measurable} if it can be written as~\cite[]{BuschBook,Ali09}
\begin{equation}
\Eax = \sum_\lambda P(a|x,\lambda) G_\lambda\quad\forall a,x,
\end{equation}
where $\{G_\lambda\}_\lambda$ is a valid POVM and $P(a|x,\lambda)$ are non-negative numbers such that $\sum_a P(a|x,\lambda)=1$ for all $x,\lambda$. Physically, joint measurability means that the statistic of each POVM in the assemblage can be obtained by classically post-processing the statistic of a parent POVM $\{G_\lambda\}_\lambda$, irrespective of the state.

Several incompatibility measures have been proposed in the literature (see Ref.~\cite[]{Designolle19} for an overview). Here, we choose 
the \emph{incompatibility robustness}~\cite[]{Haapasalo15Robustness,Uola15,Heinosaari16} defined as 
\begin{equation}
\begin{split}
\IR(\{\Eax\}):= \min\Big\{t\ \Big|\ \{(\Eax+t\cdot N_{a|x})/(1+t)\}_{a,x} \\
\text{ is jointly measurable }\Big\},
\end{split}
\end{equation}
where the minimum is take w.r.t. any arbitrary assemblage $\{N_{a|x}\}_{a,x}$. Here, $\IR$ is related to the minimum noise necessary
for $\{\Eax\}_{a,x}$ to become jointly measurable. From a quantum information perspective, $\IR$ quantifies the advantage that $\{\Eax\}_{a,x}$ provides w.r.t. jointly measurable assemblages for a certain state-discrimination task~\cite[]{Carmeli19,Skrzypczyk19,Uola19a,Takagi19PRL,Takagi19PRX,Oszmaniec2019operational,
Mori2019Operational}. Moreover, it can be efficiently computed via SDP~\cite[]{Uola15,BoydBook}:
\begin{equation}
\begin{aligned}
\IR = \min_{\{G_\lambda\}} \quad & \frac{1}{d}\sum_\lambda \tr[G_\lambda]-1\\
{\rm s.t.}\quad  & G_\lambda\succeq 0 \ \forall\lambda, \ \sum_\lambda \delta_{a,\lambda_x} G_\lambda \succeq \Eax\ \forall a,x,\ \\
&\sum_\lambda G_\lambda =  \frac{1}{d}\left(\sum_\lambda\tr[G_\lambda]\right)\cdot \openone,
\end{aligned}
\label{Eq_SDP_IR}
\end{equation}
where $\lambda:=(\lambda_1,\lambda_2,...,\lambda_{|\mathcal{X}|})$, $\lambda_i \in \mathcal{A}$, encodes the deterministic strategies.

\section{The measurement moment matrices}
As first noted by Moroder \emph{et.~al.}~\cite[]{Moroder13}, moment matrices can be interpreted as the application of a completely positive map to a (set of) positive operator(s), such as a quantum state~\cite[]{Doherty08,NPA,NPA2008,Moroder13} or steering state ensembles~\cite[]{CBLC16,CBLC18}.
Here, we define the {\it measurement moment matrices} (MMMs) by applying a completely positive map on \emph{POVMs}
\begin{equation}
\chi[\Eax]:=\sum_n K_n (\Eax\otimes\openone^{\rm B}) K_n^\dag\quad\forall a,x,
\label{Eq_CP_map}
\end{equation}
where the map is obtained by first embedding the system ${\rm A}$ in the tensor product with a second identical system ${\rm B}$, i.e., $\Eax \mapsto \Eax\otimes\openone^{\rm B}$, which is a completely positive map, and then applying the Kraus operators $K_n:{\rm AB} \rightarrow \overline{\rm AB}$ defined as ${K_n:=\sum_i \ket{i}_{\overline{\rm AB}{\rm AB}}\bra{n} (\rab)^{\frac{1}{2}}S_i}$,
with  $\{\ket{i}\}_i$ and $\{\ket{n}\}_n$ the orthonormal bases for the output space $\overline{\rm AB}$ and the input space ${\rm AB}$, respectively, and $\{S_i\}$ is a sequence of operators to be specified later.
In this way, one obtains a moment matrix
\begin{equation}
\chi_{\rab,\{S_i\}} [\Eax]= \sum_{ij}\ket{i}\bra{j}\tr\left[S_i(\Eax\otimes\openone^{\rm B})S_j^\dag\rab\right]
\label{Eq_LM}
\end{equation}
for each $a,x$. In what follows, we simply use the symbol $\chi[\Eax]$, or even $\chi$, when there is no risk of confusion. The MMM $\chi$ is a type of \emph{localizing matrix}, proposed  in the context of noncommutative polynomial optimization~\cite[]{Pironio10b}, but here we define them from the perspective of measurement effects. In particular, their formulation is independent of the standard Navascu{\'e}s-Pironio-Ac{\'i}n (NPA) moment matrix~\cite[]{NPA,NPA2008}. 

We choose the operators $\{S_i\}$ as products of POVM elements, e.g., $\{S_i\} = \{\Eax\otimes\openone^{\rm B}, \openone^{\rm A}\otimes \Eby, \Eax\otimes\Eby, \Eax\otimes (\Eby E_{b'|y'}^\text{B}), {\rm etc.} \}$, and following the convention of Ref.~\cite[]{Moroder13}: a level $\ell$ is denoted by $\{S_i^{(\ell)}\} := \openone\cup\mathcal{O}^{(1)}\cup\mathcal{O}^{(2)}\cup,...,\cup\mathcal{O}^{(\ell)}$, where $\mathcal{O}^{(\ell)}:=\{E_{a_1|x_1}^\text{A}E_{a_2|x_2}^\text{A}...E_{a_{\ell-k}|x_{\ell-k}}^\text{A}\otimes E_{b_{\ell-k+1}|y_{\ell-k+1}}^\text{B}...E_{b_{\ell}|y_{\ell}}^\text{B}\}$ is composed of all $\ell$-order products of $\Eax$'s and $\Eby$'s.
Even though the operators $\rab,\{\Eax\}_{a,x}$ and $\{\Eby\}_{b,y}$ are uncharacterized, one is still able to obtain specific entries in $\chi$, such as those corresponding to accessible statistics in a DI setting, i.e., $P(a,b|x,y)=\tr(\Eax\otimes\Eby \rab)$. Moreover, by the Neumark dilation~\cite[]{Peres90}, any POVM can be realized by a projective measurement in a higher dimensional space, implying conditions such as $0=\tr(\Eax E_{a'|x}^\text{A}\otimes \Eby)$, for $a'\neq a$, or $0=\tr(\Eax \otimes \Eby E_{b'|y}^{\rm B})$, for $b'\neq b$. Moreover, since the MMMs are obtained by applying a completely positive map on valid POVMs (see Eq.~\eqref{Eq_CP_map}), each $\chi$ is positive semidefinite by construction. It is convenient to decompose $\chi$ into the characterized parts and unknown parts~\cite[]{Moroder13}:
\begin{equation}
\begin{aligned}\label{eq:chi_fix}
\chi
&= \chi^{\rm fixed}(P) + \chi^{\rm open}(u)\\
&= \sum_{a,b,x,y} P(a,b|x,y)F_{a,b,x,y} + \sum_v u_v F_v,
\end{aligned}
\end{equation}
where all of $F_{a,b,x,y}$ and $F_v$ are symmetric matrices. The complex numbers $u_v$ represent all the uncharacterized variables.

\section{Device-independent quantification of measurement incompatibility}
Via the MMM, we are able to define, for any SDP involving effect operators, its {\it DI relaxation}, i.e., another version of the problem involving only DI assumptions. As an example, we will show below how to define the incompatibility robustness. Several other examples, such as incompatibility jointly measurable robustness, incompatibility probabilistic robustness, incompatibility random robustness, and the incompatibility weight, are described in App.~\ref{app:other_meas}. The problem in Eq.~\eqref{Eq_SDP_IR} is mapped to
\begin{equation}
\begin{aligned}
\min_{\{\chi[G_\lambda] , \chi[E_{a|x}] \}_{\lambda,a,x} } \quad & \sum_\lambda \chi[G_\lambda]_{\openone} -1\\
{\rm s.t.} \quad & \sum_\lambda \delta_{a,\lambda_x} \chi[G_\lambda]\ \succeq \chi[\Eax]\quad\forall a,x,\\
&\chi[G_\lambda]\succeq 0\quad\forall\lambda,\\
&\sum_\lambda \chi[G_\lambda] = \sum_\lambda \chi[G_\lambda]_{\openone}\cdot\chi[\openone],\\
&\sum_a \chi[\Eax]=\chi[\openone]\quad\forall x,\\
&\chi[\Eax]\succeq 0\quad\forall a,x,\\
&P(a,b|x,y) = P_{\rm obs}(a,b|x,y)\quad\forall a,b,x,y,\label{Eq_DIIR}
\end{aligned}
\end{equation}
where $\chi[G_\lambda]_{\openone}:=\tr(G_\lambda\otimes\openone^{\rm B}\rab)$. The objective function is the same as that of Eq.~\eqref{Eq_SDP_IR} due to the fact that ${\tr(\sum_\lambda G_\lambda\otimes\openone^{\rm B}\rab)=\tr[\big(\sum_\lambda\tr(G_\lambda)\big)\openone^{\rm A}\otimes\openone^{\rm B}\rab]/d} =(1/d)\sum_\lambda\tr(G_\lambda)$. The first three constrains are directly obtained from the three constraints in Eq.~\eqref{Eq_SDP_IR}. The rest are associated with, respectively, normalization of POVMs, positivity of POVMs, and the observed nonlocal correlation. The above problem is not an SDP yet, since  the third constraint in Eq.~(\ref{Eq_DIIR}) is quadratic. To tackle this problem, we relax the third constraint by keeping only the characterized terms in  $\chi[\openone]$. Namely, the relaxed constraint becomes:
$\sum_\lambda \chi[G_\lambda]^{\rm fixed} = \sum_\lambda \chi[G_\lambda]_{\openone}\cdot \chi[\openone]^{\rm fixed}$, where, with some abuse of notation (since no elements in $\chi[G_\lambda]$ are actually fixed), we mean to retain only the constraints associated with entries in $\chi[\openone]^{\rm fixed}$ as in Eq.~\eqref{eq:chi_fix}, i.e., with the observed probabilities $P_{\rm obs}(a,b|x,y)$.

\begin{equation}
\begin{aligned}
\text{ Given } \qquad &P_{\rm obs}(a,b|x,y)\\
\min_{\chi[\openone], \{\chi[G_\lambda] , \chi[E_{a|x}] \}_{\lambda,a,x} } \quad & \sum_\lambda \chi[G_\lambda]_{\openone} -1\\
{\rm s.t.} \quad & \sum_\lambda \delta_{a,\lambda_x} \chi[G_\lambda]\ \succeq \chi[\Eax]\quad\forall a,x,\\
&\chi[G_\lambda]\succeq 0\quad\forall\lambda,\\
&\sum_\lambda \chi[G_\lambda]^{\rm fixed} = \sum_\lambda \chi[G_\lambda]_{\openone}\cdot\chi[\openone]^{\rm fixed},\\
&\sum_a \chi[\Eax]=\chi[\openone]\quad\forall x,\\
&\chi[\Eax]\succeq 0\quad\forall a,x,\\
&P(a,b|x,y) = P_{\rm obs}(a,b|x,y),\\
&\forall a,b,x,y,
\end{aligned}
\label{Eq_DIIR_fix}
\end{equation}

The solution obtained above, denoted by $\DIIR$, is a lower bound on $\IR$ of the underlying measurement assemblage. In other words, it tells us the minimum degree of measurement incompatibility present
when observing a certain nonlocal correlation.

An analogous SDP can be used for bounding from below the measurements incompatibility necessary  for a given violation of Bell inequality. In this case, only the Bell value, i.e., $I(P)$, is given and not $P_{\rm obs}(a,b|x,y)$. As a consequence one simply removes entirely the third constraint in Eq.~\eqref{Eq_DIIR}, as $\chi[\openone]^{\rm fixed}$ is not characterized. Alternatively, by changing the objective function one may ask what is the maximal violation of a Bell inequality for a given value $\IR_0$ of the robustness. It can be easily shown that for each pair $(I(P), \IR_0)$ a feasible solution of one SDP is also a feasible solution of the other, hence, they characterize the same set. See App.~\ref{app:IR} for more details.

The formulation with the fixed $\IR_0$, however, turns out to more more convenient, as it removes the nonlinearity in the previous SDP. In fact, the substitution $\sum_\lambda \chi[G_\lambda]_{\openone} -1 = \IR_0$, allows us to write the third constraint of Eq.~\eqref{Eq_DIIR} as  $\sum_\lambda \chi[G_\lambda] = (\IR_0+1) \chi[\openone]$. We then have
\begin{equation}
\begin{aligned}
\text{ Given } \qquad & \IR_0 \\
\max_{\chi[\openone],\{\chi[G_\lambda] , \chi[E_{a|x}] \}_{\lambda,a,x} } \quad & I(P)\\
{\rm s.t.} \quad & \sum_\lambda \delta_{a,\lambda_x} \chi[G_\lambda]\ \succeq \chi[\Eax]\quad\forall a,x,\\
&\chi[G_\lambda]\succeq 0\quad\forall\lambda,\\
&\sum_\lambda \chi[G_\lambda] = (\IR_0+1)\chi[\openone],\\
&\sum_a \chi[\Eax]=\chi[\openone]\quad\forall x,\\
&\chi[\Eax]\succeq 0\quad\forall a,x,\\
&\sum_\lambda \chi[G_\lambda]_{\openone} = \IR_0 + 1,
\end{aligned}
\label{Eq_max_bell}
\end{equation}

We apply this method to the tilted-CHSH inequality~\cite[]{Acin12}, see the next section for a detailed explanation and Fig.~\ref{Fig_comparison_DIIR_tCHSH} for a summary of the results. What we want to highlight now, is that for the simple case analyzed in Fig.~\ref{Fig_comparison_DIIR_tCHSH}, the SDP in Eq.~\eqref{Eq_DIIR} already  provides an exact solution, despite the relaxation of the nonlinear constraint. In contrast, for the case of genuine multipartite incompatibility robustness, discussed in Sec.~\ref{sec:GMI} below, we see that different bounds arise when the same constraint is taken into account or not, see also Apps.~\ref{app:GMI} and \ref{app:Wit_GMI}.

\section{Quantification of incompatibility robustness}

\begin{figure}[t]
\emph{\includegraphics[width=8cm]{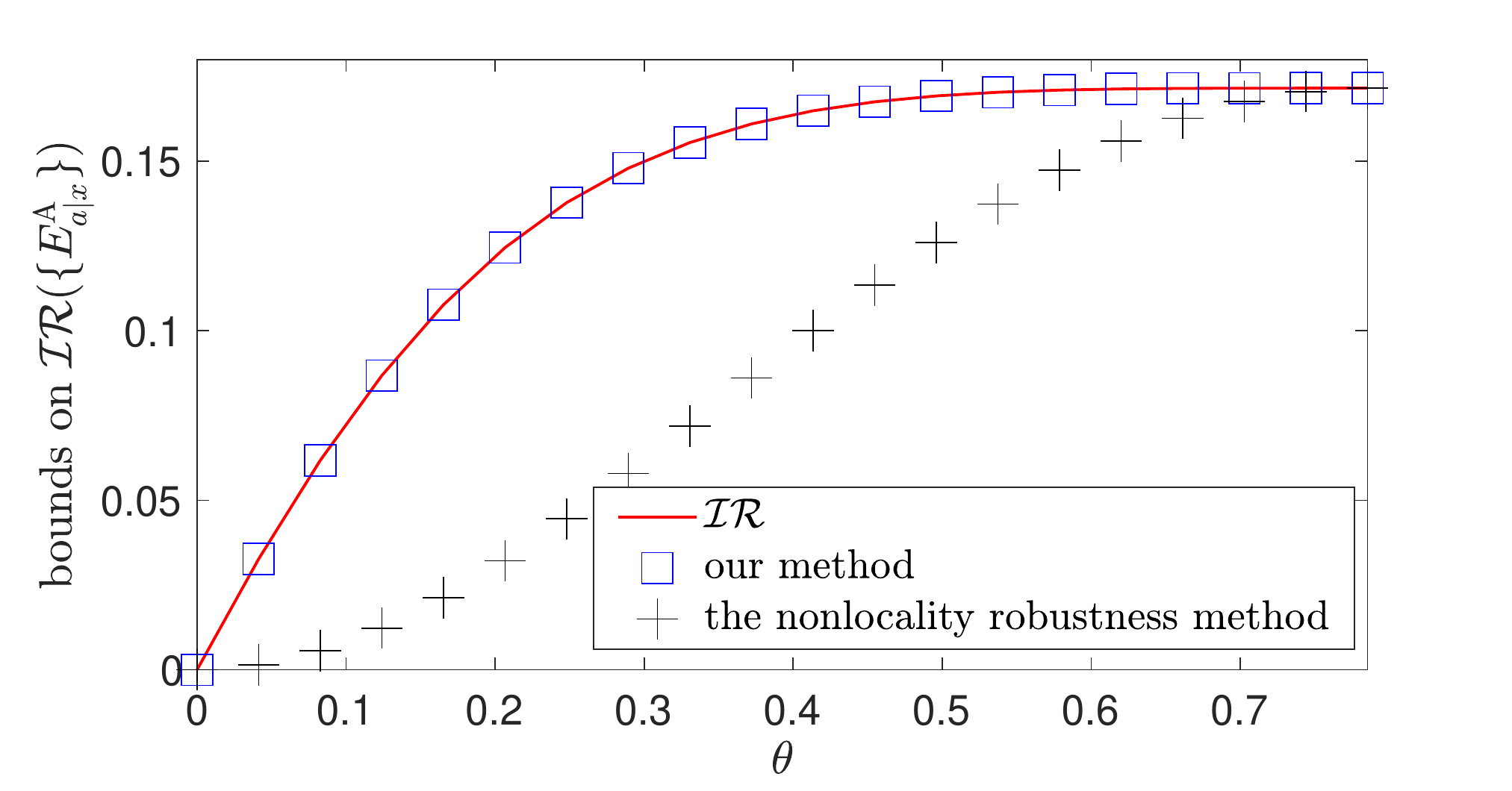} }
\caption{DI $\mathcal{IR}$ bounds with the MMM method and the nonlocal robustness (NLR). Red curve: $\mathcal{IR}$ of Bob's optimal measurements for the tilted-CHSH inequality. Blue squares: DI lower bound from the MMM method ($2$nd level of the hierarchy). Black crosses: lower bound from NLR.  }
\label{Fig_comparison_DIIR_tCHSH}
\end{figure}

As a first application of our method, we consider the simplest Bell scenario, i.e., the Clauser-Horne-Shimony-Holt (CHSH) scenario. More precisely, we consider the tilted-CHSH~\cite[]{Acin12} (see also \cite[]{Yang13,Bamps15}) inequality, parametrized by $\alpha$, namely, $
I_{\rm CHSH}^{\rm tilted}:=
\alpha\langle A_1\rangle  + \langle A_1 B_1\rangle + \langle A_1 B_2\rangle + \langle A_2 B_1\rangle - \langle A_2 B_2\rangle 
\stackrel{\mathcal{L}}{\leq} 2+\alpha$, with $\langle A_x\rangle:=P_{\rm A} (a=1|x)-P_{\rm A} (a=-1|x)$ and $\langle A_x B_y\rangle:=P(a=b|x,y)-P(a\neq b|x,y)$ being the correlators. The maximal quantum violation, $\sqrt{8+2\alpha^2}$, is achieved with two fixed Pauli measurement on Alice's side, i.e., $\hat X$ and $\hat Z$, and tilted measurements for Bob, i.e., $\cos\mu\hat{Z} + \sin\mu\hat{X}, \cos\mu\hat{Z} - \sin\mu\hat{X}$ on the partially entangled state $|\psi_\theta\rangle=\cos\theta|00\rangle + \sin\theta|11\rangle$, with $\mu=\arctan(\sin 2\theta)$ and $\theta=(1/2)\arctan(\sqrt{(4-\alpha^2)/2\alpha^2})$.

For each value of $\alpha$ one can obtain the optimal state and the optimal pair of measurements (unique up to local isometries) providing the maximal quantum violation. The value of Bob's robustness for a given $\theta$ coincides with its DI bound computed via the MMM assuming the corresponding distribution $P(a,b|x,y)$ (see Fig.~\ref{Fig_comparison_DIIR_tCHSH}). In the same figure, we also plot the DI bound of $\IR$ obtained via the  nonlocality robustness (NLR)~\cite[]{Cavalcanti16} method. The NLR method, as well as another method proposed for the DI lower bound of incompatibility, i.e., the assemblage moment matrix (AMM)~\cite[]{CBLC16,CBLC18} method, are based on the connection between steering and incompatibility~\cite[]{Uola14,Uola15,Quint14}. In contrast, the MMM relies on the construction of a moment matrix directly from the measurement operators. In App.~\ref{app:previous_methods}, we show that the AMM can be identified with a special case of a MMM. Hence, it can never provide a better bound for incompatibility. In addition, we explicitly show via the $I_{3322}$ inequality~\cite[]{Collins04}, that the MMM provides strictly better bounds.


\section{Quantification of genuine multipartite incompatibility robustness} \label{sec:GMI}
\begin{figure}[t]
\emph{\includegraphics[width=0.45\textwidth]{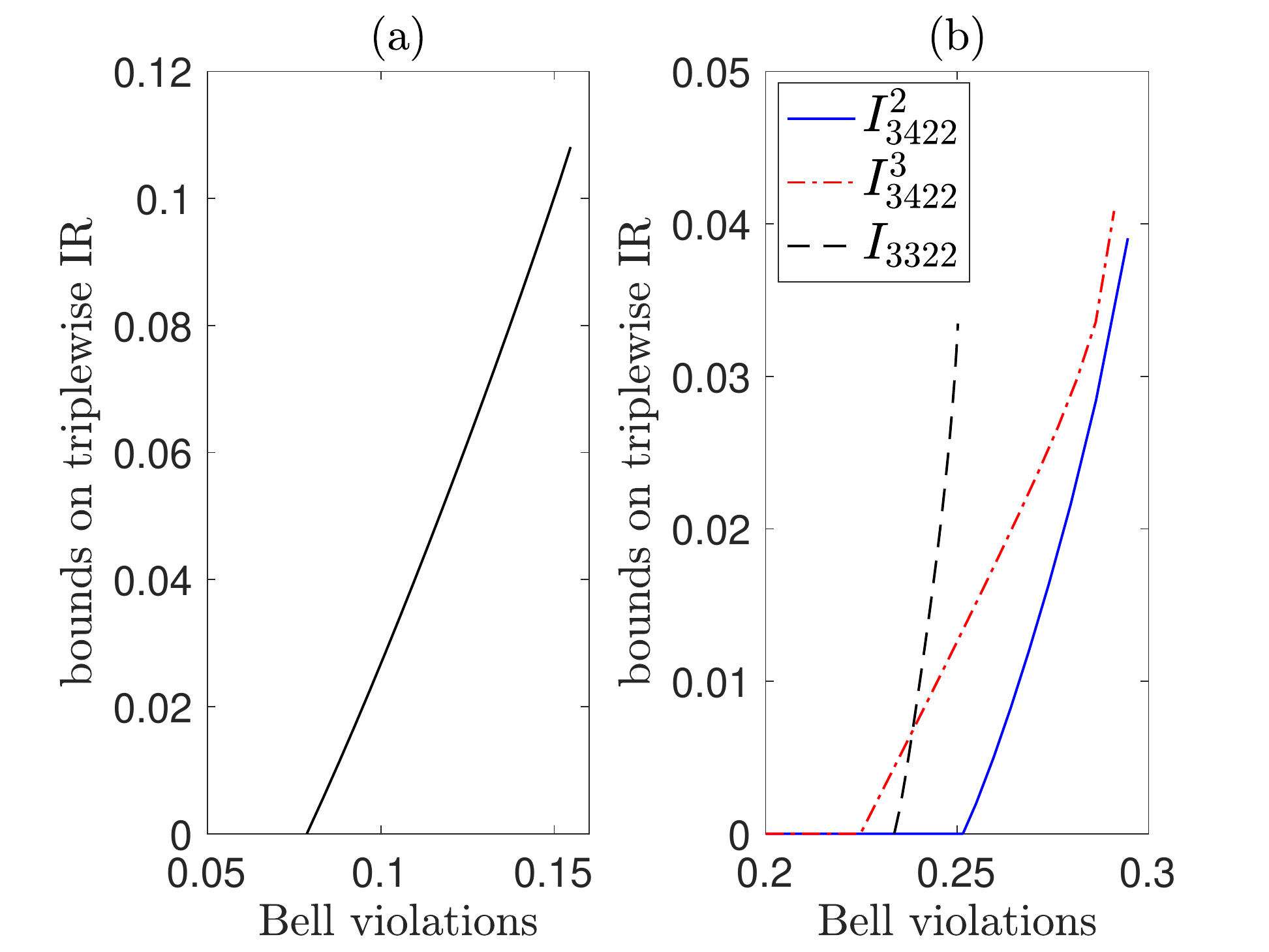} }
\caption{MMMs can also be used to compute lower bounds on genuine triplewise $\mathcal{IR}$ in a DI setting. (a) DI lower bounds on genuine triplewise $\mathcal{IR}$ in the elegant Bell scenario. (b)The black-dashed, red-dash-dotted, and blue-solid curves represent, respectively, DI lower bounds on genuine triplewise $\mathcal{IR}$ in the $I_{3322}$, $I_{3422}^{3}$, and $I_{3422}^{2}$ scenarios. The SDP carrying out the computation can be found in Eq.~\eqref{SDP_DI_GMIR_fix_rob}.
}
\label{Fig_DIGMIR}
\end{figure}

Here, we show how the MMM can be used to quantify the genuine multipartite incompatibility robustness (GMIR) recently introduced by Quintino \emph{et al}.~\cite[]{QuintinoPRL2019}. An example is provided in Fig.~\ref{Fig_DIGMIR} for different Bell inequalities. All the results presented, use the maximization of the Bell violation for a given robustness, see Eq.~\eqref{SDP_DI_GMIR_fix_rob} below. As we discuss in App.~\ref{app:GMI}, the results obtained with this method are provable better than those obtained minimizing the robustness for a given Bell violation. Finally, in addition to being able to quantify the GMIR, our method can also improve the thresholds for its detection. We compare ours with those computed in Ref.~\cite[]{QuintinoPRL2019}.

A measurement assemblage of three measurements $\{\{E_{a|x}\}_a\}_{x=1,2,3}$ is said to be 
{\it genuinely triplewise incompatible}~\cite[]{QuintinoPRL2019} if it is impossible to write it as a convex mixture of three measurement assemblages, each containing a different pair of compatible measurements \cite[]{QuintinoPRL2019}. More concretely, if there exists three assemblages $\{\{J_{a|x}^{st}\}_a\}_{x=1,2,3}$ for $(s,t) = (1,2),(1,3),(2,3)$ such that  $\{J_{a|s}^{st}\}_a$ and $\{J_{a|t}^{st}\}_a$ are jointly measurable for any pair $s,t$ and each $E_{a|x}$ can be written as 
\begin{equation}
 E_{a|x}=p_{12} J^{12}_{a|x} + p_{23} J^{23}_{a|x} + p_{13} J^{13}_{a|x} 
\end{equation}
for some probabilities $p_{12}$, $p_{23}$, and $p_{13}$ that respect $p_{12}+p_{23}+p_{13}=1$, we will say that $\{\{E_{a|x}\}_a\}_{x=1,2,3}$  are {\it not} genuine triplewise incompatible.

This condition can be written in a SDP form (see Ref.~\cite{QuintinoPRL2019} and App.~\ref{app:GMI} for a brief self-contained summary), which leads to a SDP formulation of the robustness as
\begin{align}\label{SDP_GMIR}
 \text{Given }& \{E_{a|x}\}_{a,x}, \nonumber \\ 
 \text{and variables } &\{ G^{12}_\lambda , G^{13}_\lambda , G^{23}_\lambda \}_\lambda, \nonumber
 \{ J_{a|3}^{12},J_{a|2}^{13},J_{a|1}^{23} \}_{a} ,  \nonumber\\
\min\ & \frac{1}{d}\sum_\lambda \tr[G^{12}_\lambda + G^{13}_\lambda + G^{23}_\lambda ] - 1\\
\text{s.t. } & G^{st}_{\lambda} \succeq 0\ \forall \lambda,\ \sum_\lambda G^{st}_\lambda = \frac{\openone}{d} \sum_\lambda \tr[G^{st}_\lambda]\nonumber\\
& \text{ for } (s,t)=(1,2),(1,3),(2,3); \nonumber \\
& J^{st}_{a|x} \succeq 0, \ \forall a,\  \sum_a J^{st}_{a|x} = \sum_\lambda G^{st}_\lambda \text{ and }\nonumber \\
& \sum_\lambda \delta_{a,\lambda_x} (G^{sx}_\lambda + G^{tx}_\lambda) + J^{st}_{a|x} \succeq E_{a|x},\nonumber\\
& \text{ for } (s,t,x)=(1,2,3),(1,3,2),(2,3,1); \nonumber 
\end{align}

Applying the same argument as the one for the standard incompatibility robustness above SDP can have a DI relaxation via moment matrices
\begin{align}\label{SDP_DI_GMIR}
\text{Given}\qquad & P_{\rm obs}(a,b|x,y), \text{ and }\nonumber \\
\text{variables}\ & \{\chi[E_{a|x}]\}_{a,x}, \text{ and }\{\chi[G^{st}_\lambda]\}_\lambda, \{\chi[J^{st}_{a|x}]\}_{a},  \nonumber\\
&\text{ for } (s,t,x) = (1,2,3), (1,3,2),(2,3,1); \nonumber\\
\min\ & \sum_\lambda \chi[G^{12}_\lambda]_{\openone} + \chi[G^{13}_\lambda]_{\openone} + \chi[G^{23}_\lambda ]_{\openone} - 1 \nonumber \\
\text{s.t. } & \chi[G^{st}_{\lambda}] \succeq 0\ \forall \lambda, (s,t)=(1,2),(1,3),(2,3); \nonumber \\ 
&\sum_{\lambda} \chi[G^{st}_\lambda]^{\rm fixed} =  \chi[\openone]^{\rm fixed} \sum_{\lambda} \chi[G^{st}_\lambda]_{\openone}\  \nonumber\\
& \text{ for } (s,t)=(1,2),(1,3),(2,3); \nonumber \\
& \chi[J^{st}_{a|x}] \succeq 0, \ \forall a,\  \sum_a \chi[J^{st}_{a|x}] = \sum_\lambda \chi[G^{st}_\lambda] \text{ and }\nonumber \\
& \sum_\lambda \delta_{a,\lambda_x} (\chi[G^{sx}_\lambda] + \chi[G^{tx}_\lambda]) + \chi[J^{st}_{a|x}] \succeq \chi[E_{a|x}],\nonumber\\
& \text{ for } (s,t,x)=(1,2,3),(1,3,2),(2,3,1); \nonumber \\
& \chi[E_{a|x}] \succeq 0, \text{ for all } a,x, \nonumber \\
& \sum_a \chi[E_{a|x}] = \chi[\openone], \text{ for all } x, \nonumber \\
&P(a,b|x,y) = P_{\rm obs}(a,b|x,y).
\end{align}

Again, one can compute the maximum of a Bell inequality $I(P)$ for a given robustness $\IR_0$ as
\begin{align}\label{SDP_DI_GMIR_fix_rob}
\text{Given} \ & \ \IR_0,\nonumber \text{ and }\\
\text{variables}\ & \{\chi[E_{a|x}]\}_{a,x},\{\chi[G^{st}_\lambda]\}_\lambda, \{\chi[J^{st}_{a|x}]\}_{a},  \nonumber\\
&\text{ for } (s,t,x) = (1,2,3), (1,3,2),(2,3,1), \nonumber\\
\max\ & I(P) \nonumber \\
\text{s.t. } & \chi[G^{st}_{\lambda}] \succeq 0\ \forall \lambda, (s,t)=(1,2),(1,3),(2,3); \nonumber \\ 
&\sum_{\lambda,(s,t)} \chi[G^{st}_\lambda] =  \chi[\openone] (\IR_0+1)\ \text{ and } \nonumber\\
&\sum_{\lambda,(s,t)} \chi[G^{st}_\lambda]_{\openone} =  (\IR_0+1),\nonumber \\
& \text{ with sum over } (s,t)=(1,2),(1,3),(2,3); \nonumber \\
& \chi[J^{st}_{a|x}] \succeq 0, \ \forall a,\  \sum_a \chi[J^{st}_{a|x}] = \sum_\lambda \chi[G^{st}_\lambda] \text{ and }\nonumber \\
& \sum_\lambda \delta_{a,\lambda_x} (\chi[G^{sx}_\lambda] + \chi[G^{tx}_\lambda]) + \chi[J^{st}_{a|x}] \succeq \chi[E_{a|x}],\nonumber\\
& \text{ for } (s,t,x)=(1,2,3),(1,3,2),(2,3,1); \nonumber \\
& \chi[E_{a|x}] \succeq 0, \text{ for all } a,x, \nonumber \\
& \sum_a \chi[E_{a|x}] = \chi[\openone], \text{ for all } x.
\end{align}	


\begin{table}[t]
\centering
\begin{tabular}{|c|c|c|c|}
\hline
Bell inequality & Tab.I of \cite[]{QuintinoPRL2019} & \quad NPA+comm.\quad\phantom{.} & MMM$(\ell=2)$  \\ \hline \hline
$I_{\rm E}$     & $0.0786$ & $0.0786$ & $0.0786$    \\ \hline
$I_{3422}^2$ & $0.2768$  &  $0.2647$  & $0.2515$    \\ \hline
$I_{3422}^3$ & $0.2615$ & $0.2247$ & $0.2247$    \\ \hline
$I_{3322}$     & $0.2487$  &  $0.2387$  & $0.2335$    \\ \hline
\end{tabular}
\caption{Comparison of the thresholds for Bell-inequality violations able to certify genuine tripartite incompatibility.  Our method always performs  better than the one based on nonlocality arguments (the set $L^2_{\rm conv}$ in \cite[]{QuintinoPRL2019}) and better or equal than the NPA hierarchy with additional commutativity constraints (the set $Q_{2{\rm conv}_{JM}}$ defined in \cite[]{QuintinoPRL2019}, see App.~\ref{app:Wit_GMI} for details). We recall that the bound for 
$I_{\rm E}$ is tight, as proven in \cite[]{QuintinoPRL2019}.}\label{TB1}
\end{table}


As we mention above, in this case one can show that the problem in Eq.~\eqref{SDP_DI_GMIR_fix_rob}, namely, the maximization of the Bell violation for a given robustness, provides better bounds than the inverse problem, namely, the minimization of the robustness for a given Bell violation. This is due to the possibility of removing the nonlinear constraint present in the intermediate formulation.   More details can be found in App.~\ref{app:GMI}.

In addition to the quantitative results plotted in Fig.~\ref{Fig_DIGMIR}, our method is also able to improve the numerical thresholds for the detection of GMI previously found in Ref.~\cite{QuintinoPRL2019}, see Tab.~\ref{TB1} and App.~\ref{app:Wit_GMI} for more details.

\section{Semi-device-independent approach and projective measurements} 
Another advantage of our method is that it admits a direct extension to semi-device-independent (SDI) characterization of incompatibility. This can be achieved by employing ideas from the Navascu\'es-Vertesi hierarchy~\cite[]{Navascues15prl}, which generalizes the NPA hierarchy and aims to bound the set of finite dimensional quantum correlations. The key idea of this generalization comes from the fact that moment matrices generated by states and measurements of a given Hilbert space dimension $d$ span only a subspace $\mathcal{S}_d$ of the whole space of moment matrices. One can then try to add the corresponding constraint to the problem in Eq.~(\ref{Eq_DIIR}). In practice, this is achieved by generating a basis of random moment matrices (e.g., by means of the Gram–Schmidt process) by sampling states and measurements of a given dimension.

In contrast to the DI approach, in which all POVMs can be dilated to projective measurements by increasing the system's dimension, in the SDI approach one can additionally impose the constraint that the measurements $\Eax$ are projective. 

We tried several approaches to the SDI quantification of measurement incompatibility, with and without the assumption of projective measurements. A few of them, which work in the case of Bell inequalities~\cite[]{Navascues15prl}, do not generalize to the case of incompatibility quantification, either for fundamental reasons or because they fail to provide an improvement in the numerical results for the cases analyzed. A summary of these approaches is given in  App.~\ref{app:POVM_dil}.

The most successful approach is the one in which dimension constraints are imposed by requiring that the observed probabilities are generated by a system of bounded dimension. In this case, since we are restricting ourselves to dichotomic measurements, we can use the fact that correlations generated by projections are extremal. Let us denote by $\Gamma\in \mathcal{S}_d$ the moment matrix generated via the NV method, assuming that the measurements are projective, and $\Gamma_{P(a,b|x,y)}$ the matrix entry corresponding to the observed probability $P(a,b|x,y)$. The SDP 
for the computation of the minimal robustness associated to a violation $K$ of a Bell inequality $I(P)$, can be written as
\begin{equation}
\begin{aligned}
\text{ Given } \qquad &K\\
\min_{\{\chi[G_\lambda] , \chi[E_{a|x}] \}_{\lambda,a,x} } \quad & \sum_\lambda \chi[G_\lambda]_{\openone} -1\\
{\rm s.t.} \quad & \sum_\lambda \delta_{a,\lambda_x} \chi[G_\lambda]\ \succeq \chi[\Eax]\quad\forall a,x,\\
&\chi[G_\lambda]\succeq 0\quad\forall\lambda,\\
&\sum_a \chi[\Eax]=\chi[\openone]\quad\forall x,\\
&\chi[\Eax]\succeq 0\quad\forall a,x,\\
&I(P) = K,\\
&\Gamma \in \mathcal{S}_d, \quad \Gamma \succeq 0,\\
&P(a,b|x,y) = \Gamma_{P(a,b|x,y)}\quad\forall a,b,x,y,
\end{aligned}
\label{Eq_DIIR_SDI_prob}
\end{equation}
where $P(a,b|x,y) $ denotes the entries in the MMM $\{\chi[\Eax]\}_{a,x}$, in the usual DI approach, corresponding to the probability $P(a,b|x,y)$, and $\Gamma$, as discussed above, is generated by sampling moment matrices generated with dichotomic projective measurements in dimension $d$.

Equivalently, one can fix the robustness $\IR_0$  and maximize the Bell inequality violation, as follows 
\begin{equation}
\begin{aligned}
\text{ Given } \qquad &\IR_0\\
\max_{\{\chi[G_\lambda], \chi[E_{a|x}] \}_{\lambda,a,x} }  &I(P)\\
{\rm s.t.} \quad & \sum_\lambda \delta_{a,\lambda_x} \chi[G_\lambda]\ \succeq \chi[\Eax]\quad\forall a,x,\\
&\chi[G_\lambda]\succeq 0\quad\forall\lambda,\\
&\sum_\lambda \chi[G_\lambda] = (\IR_0+1)\chi[\openone],\\
&\sum_\lambda \chi[G_\lambda]_{\openone} = \IR_0 +1,\\
&\sum_a \chi[\Eax]=\chi[\openone]\quad\forall x,\\
&\chi[\Eax]\succeq 0\quad\forall a,x,\\
&\Gamma \in \mathcal{S}_d, \quad \Gamma \succeq 0,\\
&P(a,b|x,y) = \Gamma_{P(a,b|x,y)}\quad\forall a,b,x,y,
\end{aligned}
\label{Eq_DIIR_SDI_prob2}
\end{equation}
with the same use of notation as above. 

\begin{figure}[t]
\includegraphics[width=8cm]{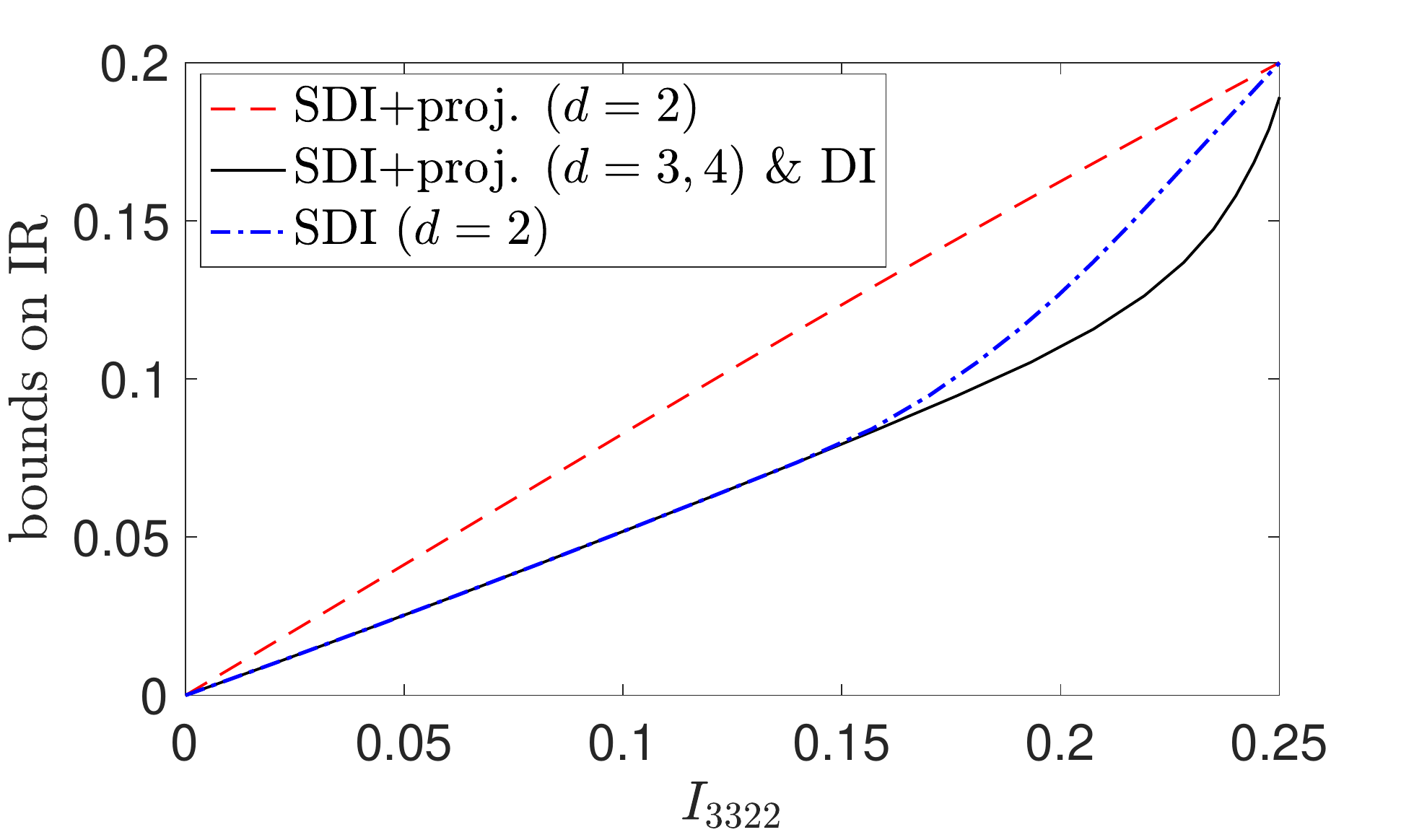} 
\caption{Quantification of incompatibility robustness for given violation of $I_{3322}$. Blue dash-dotted line: SDI approach for $d=2$. Black solid line: DI approach. A similar curve (with a difference of the order of $10^{-5}$) corresponds to the SDI calculation with the additional assumption of projective measurements for $d=3,4$ (identical up to numerical precision). Red dashed line: SDI approach with additional assumption of projective measurements for $d=2$.
}
\label{Fig_semiDI_DI_IR}
\end{figure}

In order to compare the different methods, we computed different lower bounds on the incompatibility robustness for a given violation of the $I_{3322}$ inequality. First, 
we tried the dilation method presented in Eq.~\eqref{eq:dilation_sdi} (in App.~\ref{app:POVM_dil}) for $d=2$, which gave no improvement over the standard $DI$ approach. In contrast, the SDP in Eq.~\eqref{Eq_DIIR_SDI_prob}, for $d=2$, provided a substantially improved lower bound on the robustness, with respect to the DI case. In addition, we also compare the SDI approach with the one where the additional condition of projective measurements is assumed. With the assumption of projective measurements, we were able to obtain a substantially improved bound for the case of $d=2$, whereas the case $d=3$ and $d=4$, which provided identical bounds up to numerical precision, improved only slightly the $DI$ bound, with a difference of the order of $10^{-5}$. All the corresponding curves are plotted in Fig.~\ref{Fig_semiDI_DI_IR}. All calculations were performed with the $2+$ level of the hierarchy (i.e., the second level plus additional terms) corresponding to a moment matrix of size $34\times 34$. Moreover, notice also how the curve for the case SDI plus projective measurement is concave. This is not in contradiction with our definition of the SDP: A convex mixture $\lambda \chi_1 + (1 - \lambda)\chi_2$, of a solution $\chi_1$ for $\IR_1$ and $\chi_2$ for $\IR_2$, does not necessarily provide a valid solution for the robustness $\lambda \IR_1 + (1-\lambda)\IR_2$, because both $\chi$ and $\IR$ enter the constraint $\sum_\lambda \chi[G_\lambda] = (\IR_0+1)\chi[\openone]$ in a nonlinear way.

Finally, an analogous procedure allows us to extend the MMM to another typical SDI scenario, namely the {\it prepare-and-measure} scenario. More details can be found in App.~\ref{app:SDI}.


\section{Conclusions and outlook}
We proposed a framework, the MMM,  to quantify the degree of (several notions of) measurement incompatibility in a DI manner. The main idea behind our method is to construct moment matrices by applying a completely positive map on POVMs. 
 Due to the operational characterization of the incompatibility robustness~\cite[]{Carmeli19,Skrzypczyk19,Uola19a,Takagi19PRL,Takagi19PRX,Oszmaniec2019operational,Mori2019Operational}, our result also bounds, in a DI scheme, the usefulness of a set of POVMs in the problem of quantum state discrimination. In contrast to previous DI bounds of incompatibility in Refs.~\cite[]{Cavalcanti16,CBLC16}, our method does not rely on any concept of steering, but provides a direct interpretation of the moment matrix as a completely positive mapping of the measurement operators. Our MMM method is shown to outperform both methods in the quantification of incompatibility in simple examples, and we rigorously proven that it always performs better or equal to the method in Ref.~\cite[]{CBLC16}. Moreover, the MMM method provides a DI bound of the genuine multipartite incompatibility, a recently introduced notion, for which no DI quantifier was known so far, and it improves the known thresholds for the its detection. Finally, given its generality our method is straightforwardly adaptable to include additional constraints such as the system dimension (semi-DI approach), the assumption of projective measurements, and it is applicable to the prepare-and-measure scenario  (see the discussion in App.~\ref{app:SDI}). 

We leave as an open problem to determine the convergence of the proposed hierarchy. Since we could not give either a positive or negative answer to this question, we used the term ``relaxation'' for the optimization problems throughout the text. However, we would like to point out that at least in the case of tilted CHSH, for which an analytical solution is known, our method recovers the exact relation between the incompatibility robustness and nonlocality  (see Fig. \ref{Fig_comparison_DIIR_tCHSH}).
 
As a future research direction, we would like to investigate the connection between DI and SDI quantifier of incompatibility and self-testing of measurements (see Ref.~\cite[]{Kaniewski19} for a related approach). In fact, in Ref.~\cite[]{Designolle19}, the authors showed that for the incompatibility robustness, pairs of measurements associated with mutually unbiased bases (MUBs) are the most incompatible in any dimension, even if it is not proven that they are the only ones. In the CHSH scenario, our calculation showed that $\DIIR$ saturates $\IR$ of a pair of qubit-measurements  corresponding to the MUB for the maximal quantum violation of the CHSH inequality (Ref.~\cite[]{CBLC16} also saurates this bound). For high dimensional cases, one can use the family of Bell inequalities in Ref.~\cite[]{Tavakoli19} to compute $\DIIR$. Due to the limitation of our computational capacity, we leave this issue for the potential future research. This intuition is further strengthened by the work of Ref.~\cite[]{Chen2020Robust}, which showed that the assemblage moment matrices proposed in Ref.~\cite[]{CBLC16} can be used to self-test state assemblages. Therefore it is natural to ask if the MMMs can be analogously used to self-test quantum measurements. 
Finally, a possible further extension of this work is in the direction of the SDI characterization of incompatibility in the prepare-and-measure scenario. In fact, it is believed that incompatible measurements are necessary for quantum advantage in the so-called random access codes~\cite[]{Carmeli_2020}.

\acknowledgements The authors thank Miguel Navascu\'es and Roope Uola for useful discussions. SLC acknowledges the support of the Ministry of Science and Technology, Taiwan (MOST Grants No. 109-2811-M-006-509). NM acknowledge partial support from the Foundation for Polish Science (IRAP project, ICTQT, contract no. 2018/MAB/5, co-financed by EU within Smart Growth Operational Programme). CB acknowledges the support of the Austrian Science Fund (FWF) through the projects  ZK 3 (Zukunftskolleg), and F7113 (BeyondC). YNC acknowledges the support of the Ministry of Science and Technology, Taiwan (MOST Grants No. 107-2628-M-006-002-MY3 and No. MOST 109-2627-M-006-004), and the U.S. Army Research Office (ARO Grant No. W911NF-19-1-0081).

\appendix



\section{Different measures of global incompatibility}\label{app:other_meas}

In this section, we consider other measures of incompatibility and explicitly write down their DI quantifications in the SDP form. There are robustness-based measures:
\begin{equation}
\begin{aligned}
\IR^i:= &\min\quad t\\
{\rm s.t.}\quad &\left\{\frac{\Eax+t\cdot N_{a|x}}{1+t}\right\}_{a,x}{\rm ~is~ jointly~ measurable},
\end{aligned}
\end{equation}
where the noisy models $\{N_{a|x}\}_{a,x}$ satisfy different constraints (see, e.g., \cite[]{Designolle19}) and each type of models is denoted by superindices $i$. The last measure we consider is the incompatibility weight~\cite[]{Pusey15}. For the simplicity of formulation of the following SDPs we will not write explicitly the variables of optimization. Instead, we specify the input to each SDP next to ``Given''.

\subsection{The incompatibility jointly measurable robustness}\label{SecApp_IRJ}
The noisy assemblage $\{N_{a|x}\}_{a,x}$ for the incompatibility jointly measurable robustness $\IRJ$~\cite[]{Cavalcanti16} admits a jointly measurable model. As such, $\IRJ$ can be computed via the following SDP:
\begin{equation}
\begin{aligned}
\text{ Given } \qquad &\{E_{a|x}\}_{a,x}\\
\min\quad &\frac{1}{d}\sum_\lambda \tr[H_\lambda]\\
\text{s.t.}\quad & \Eax = \sum_\lambda \delta_{a,\lambda_x}(G_\lambda-H_\lambda)\quad\forall a,x,\\
& G_\lambda\succeq 0,\quad H_\lambda\succeq 0\quad\forall\lambda,\\
& \frac{1}{d}\sum_\lambda \tr[H_\lambda] = \left(\frac{1}{d}\sum_\lambda \tr[G_\lambda]\right) - 1,\\
&\sum_\lambda H_\lambda = \left(\frac{1}{d}\sum_\lambda \tr[H_\lambda]\right)\cdot\openone,\\
&\sum_\lambda G_\lambda = \left(\frac{1}{d}\sum_\lambda \tr[G_\lambda]\right)\cdot\openone.
\end{aligned}
\end{equation}
By applying the MMM and removing the constrains containing quadratic free variables, the solution of the following SDP gives a lower bound on $\IRJ$:\footnote{We omit the description of ``for all indices'' such as ``$\forall~\lambda$'' and ``$\forall~a,x$'' when there is no risk of confusion.}
\begin{equation}
\begin{aligned}
\text{ Given } \qquad &P_{\rm obs}(a,b|x,y)\\
\min \quad & \sum_\lambda \chi[H_\lambda]_{\openone}\\
{\rm s.t.} \quad &  \chi[\Eax] = \sum_\lambda \delta_{a,\lambda_x}\Big(\chi[G_\lambda]-\chi[H_\lambda]\Big)\\
&\chi[G_\lambda]\succeq 0,\quad\chi[H_\lambda]\succeq 0,\ \ \forall \lambda,\\
& \sum_\lambda\chi[H_\lambda]_{\openone} =\left( \sum_\lambda\chi[G_\lambda]_{\openone}\right) - 1,\\
&\sum_\lambda \chi[G_\lambda]^{\rm fixed} = \left(\sum_\lambda \chi[G_\lambda]_{\openone}\right)\cdot\chi[\openone]^{\rm fixed},\\
&\sum_\lambda \chi[H_\lambda]^{\rm fixed} = \left(\sum_\lambda \chi[H_\lambda]_{\openone}\right)\cdot\chi[\openone]^{\rm fixed},\\
&\sum_a \chi[\Eax]=\chi[\openone], \ \forall x\\
&\chi[\Eax]\succeq 0,\ \forall a,x,\\
&P(a,b|x,y) = P_{\rm obs}(a,b|x,y), \ \forall a,b,x,y,
\end{aligned}
\label{EqApp_DIIRJ}
\end{equation}
where $\chi[G_\lambda]^{\rm fixed}$ and $\chi[H_\lambda]^{\rm fixed}$ in the fourth and fifth constraints respectively denote, as in the main text, $\chi[G_\lambda]$ and $\chi[H_\lambda]$ retaining entries whose indices correspond to non-vanishing terms in $\chi[\openone]^{\rm fixed}$.

\subsection{The incompatibility probabilistic robustness}\label{SecApp_IRP}
The noisy model for the incompatibility probabilistic robustness $\IRP$~\cite[]{Heinosaari14} is defined as $N_{a|x} = p(a|x)\cdot\openone$ for all $a,x,$ with real numbers $p(a|x)$ satisfying $p(a|x)\geq 0$ for all $a,x,$ and $\sum_a p(a|x)=1$ for all $x$. The associated SDP can then written as:
\begin{equation}
\begin{aligned}
\text{ Given } \qquad &\{E_{a|x}\}_{a,x}\\
\min\quad &\left(\frac{1}{d}\sum_\lambda \tr[G_\lambda]\right) - 1\\
\text{s.t.}\quad & \Eax = \sum_\lambda \delta_{a,\lambda_x}G_\lambda - q(a|x)\cdot\openone\quad\forall a,x,\\
& G_\lambda\succeq 0\quad\forall \lambda,\quad q(a|x)\geq 0\quad\forall a,x,\\
& \sum_a q(a|x) = \left(\frac{1}{d}\sum_\lambda \tr[G_\lambda]\right) - 1,\\
&\sum_\lambda G_\lambda = \left(\frac{1}{d}\sum_\lambda \tr[G_\lambda]\right)\cdot\openone.
\end{aligned}
\end{equation}
By applying the MMM, a DI lower bound can be computed via the following SDP:
\begin{equation}
\begin{aligned}
\text{ Given } \qquad &P_{\rm obs}(a,b|x,y)\\
\min \quad & \left(\sum_\lambda \chi[G_\lambda]_{\openone}\right) - 1\\
{\rm s.t.} \quad
&  \chi[\Eax]^{\rm fixed} = \sum_\lambda \delta_{a,\lambda_x}\chi[G_\lambda]^{\rm fixed}-q(a|x)\chi[\openone]^{\rm fixed},\\
&\chi[G_\lambda]\succeq 0,\ \forall \lambda, \quad q(a|x)\geq 0,\  \forall a,x,\\
&  \sum_a q(a|x) = \left(\sum_\lambda\chi[G_\lambda]_{\openone}\right) - 1\\
&\sum_\lambda \chi[G_\lambda]^{\rm fixed} = \left(\sum_\lambda \chi[G_\lambda]_{\openone}\right)\cdot\chi[\openone]^{\rm fixed},\\
&\sum_a \chi[\Eax]=\chi[\openone^{\rm A}],\ \forall x,\\
&\chi[\Eax]\succeq 0, \ \forall a,x,\\
&P(a,b|x,y) = P_{\rm obs}(a,b|x,y) \ \forall a,b,x,y.
\end{aligned}
\label{EqApp_DIIRP}
\end{equation}

\subsection{The incompatibility random robustness}\label{SecApp_IRR}
The final robustness-based measure is the incompatibility random robustness $\IRR$~\cite[]{Heinosaari15,Uola15}, where the noisy assemblage is composed of the white noise: $N_{a|x} = (1/|\mathcal{A}|)\cdot\openone$. As a result, the corresponding SDP is given by
\begin{equation}
\begin{aligned}
\text{ Given } \qquad &\{E_{a|x}\}_{a,x}\\
\min\quad &\left(\frac{1}{d}\sum_\lambda \tr[G_\lambda]\right) - 1\\
\text{s.t.}\quad & \Eax = \sum_\lambda \delta_{a,\lambda_x}G_\lambda - \frac{1}{|\mathcal{A}|}\left(\Big(\frac{1}{d}\sum_\lambda \tr[G_\lambda]\Big) - 1\right)\cdot\openone,\\
& G_\lambda\succeq 0,\ \forall \lambda,\\
&\sum_\lambda G_\lambda = \left(\frac{1}{d}\sum_\lambda \tr[G_\lambda]\right)\cdot\openone.
\end{aligned}
\end{equation}
With the same technique, a DI lower bound on $\IRR$ can be computed via the following SDP:
\begin{equation}
\begin{aligned}
\text{ Given } \qquad &P_{\rm obs}(a,b|x,y)\\
\min \quad & \left(\sum_\lambda \chi[G_\lambda]_{\openone}\right) - 1\\
{\rm s.t.} \quad
&  \chi[\Eax]^{\rm fixed} = \sum_\lambda \delta_{a,\lambda_x}\chi[G_\lambda]^{\rm fixed}\\
&\quad\quad\quad\quad\quad\quad- \frac{1}{|\mathcal{A}|}\left(\sum_\lambda \chi[G_\lambda]_{\openone} - 1\right)\cdot\chi[\openone]^{\rm fixed},\\
&\sum_\lambda \chi[G_\lambda]^{\rm fixed} = \left(\sum_\lambda \chi[G_\lambda]_{\openone}\right)\cdot\chi[\openone]^{\rm fixed},\\
&\sum_a \chi[\Eax]=\chi[\openone^{\rm A}], \ \forall x,\\
&\chi[G_\lambda]\succeq 0, \ \forall \lambda,\\
& \chi[\Eax]\succeq 0, \ \forall a,x,\\
&P(a,b|x,y) = P_{\rm obs}(a,b|x,y), \ \forall a,b,x,y.
\end{aligned}
\label{EqApp_DIIRP2}
\end{equation}

\subsection{The incompatibility weight}
The last measure of incompatibility we consider is the incompatibility weight $\IW$~\cite[]{Pusey15}. Consider that one decomposes $\Eax$ into $\Eax = t O_{a|x} + (1-t) N_{a|x}$, where $\{O_{a|x}\}_{a,x}$ is any valid quantum measurement assemblage and $\{N_{a|x}\}_{a,x}$ is a jointly measurable measurement assemblage. $\IW$ is defined as the minimum ratio of $O_{a|x}$, i.e., the minimum value of $t$, required to decompose $\Eax$. Consequently, $\IW$ can be computed via the following SDP:
\begin{equation}
\begin{aligned}
\text{ Given } \qquad &\{E_{a|x}\}_{a,x}\\
\min \quad & 1 - \frac{1}{d}\sum_\lambda \tr[G_\lambda]\\
{\rm s.t.} \quad & \Eax \succeq \sum_\lambda \delta_{a,\lambda_x} G_\lambda \quad\forall a,x,\\
&G_\lambda\succeq 0\quad\forall\lambda,\\
&\sum_\lambda G_\lambda =  \frac{1}{d}\left(\sum_\lambda\tr[G_\lambda]\right)\cdot \openone,
\end{aligned}
\label{EqApp_SDP_IW}
\end{equation}
Following the same procedure as in the previous sections, we obtain the following SDP, which can be used to compute a DI lower bound on $\IW$:
\begin{equation}
\begin{aligned}
\text{ Given } \qquad &P_{\rm obs}(a,b|x,y)\\
\min \quad & 1-  \sum_\lambda \chi[G_\lambda]_{\openone}\\
{\rm s.t.} \quad &  \chi[\Eax]\  \succeq \sum_\lambda \delta_{a,\lambda_x} \chi[G_\lambda],\  \\
&\chi[G_\lambda]\succeq 0,\ \forall \lambda,\\
&\sum_\lambda \chi[G_\lambda]^{\rm fixed} = \sum_\lambda \chi[G_\lambda]_{\openone}\cdot\chi[\openone]^{\rm fixed},\\
&\sum_a \chi[\Eax]=\chi[\openone],\ \forall x,\\
&\chi[\Eax]\succeq 0, \ \forall a,x,\\
&P(a,b|x,y) = P_{\rm obs}(a,b|x,y), \ \forall a,b,x,y.
\end{aligned}
\label{EqApp_DIIW}
\end{equation}

Note that all of above SDPs that compute DI lower bounds on the degree of incompatibility require the detailed information about the observed correlation $\{P_{\rm obs}(a,b|x,y)\}_{a,b,x,y}$. If one is merely concerned with a Bell inequality violation without the specific characterization of $\{P_{\rm obs}(a,b|x,y)\}_{a,b,x,y}$, the constrains containing $\chi[\openone]^{\rm fixed}$ have to be fully removed.


\section{Different constraints on incompatibility robustness}\label{app:IR}


As we discussed in the main text, different relaxations of the following problem exist
\begin{equation}
\begin{aligned}
\min_{\{\chi[G_\lambda] , \chi[E_{a|x}] \}_{\lambda,a,x} } \quad & \sum_\lambda \chi[G_\lambda]_{\openone} -1\\
{\rm s.t.} \quad & \sum_\lambda \delta_{a,\lambda_x} \chi[G_\lambda]\ \succeq \chi[\Eax]\quad\forall a,x,\\
&\chi[G_\lambda]\succeq 0\quad\forall\lambda,\\
&\sum_\lambda \chi[G_\lambda] = \sum_\lambda \chi[G_\lambda]_{\openone}\cdot\chi[\openone],\\
&\sum_a \chi[\Eax]=\chi[\openone]\quad\forall x,\\
&\chi[\Eax]\succeq 0\quad\forall a,x,\\
&P(a,b|x,y) = P_{\rm obs}(a,b|x,y)\quad\forall a,b,x,y,\label{Eq_DIIR_app}
\end{aligned}
\end{equation}
which are necessary to remove the nonlinear constraint: $\sum_\lambda \chi[G_\lambda] = \sum_\lambda \chi[G_\lambda]_{\openone}\cdot\chi[\openone]$. Moreover, the problem in Eq.~\eqref{Eq_DIIR_app} assume the knowledge of the full distribution of probabilities  $\{P_{\rm obs}(a,b|x,y)\}_{a,b,x,y}$, whereas in some cases, we may want to estimate the robustness simply from the violation of a Bell inequality. 

In this case, we want to characterize the set of all possible pairs $(\IR, I(P))$, where $\IR$ represents the incompatibility robustness and $I(P)$ the value of some Bell expression. Notice that, even if $I(P)$ is evaluated on a probability distribution $P$, we are not assuming that such $P$ is directly accessible,  the parameter in our problem is only the value of the Bell expression.

The set of valid $(\IR, I(P))$ can be defined by the following SDP (feasibility problem):
\begin{equation}
\begin{aligned}
\text{ Given } \qquad & \IR_0,\ I(P_0) \\
\text{ find} \qquad  & \chi[\openone], \chi[G_\lambda] , \chi[E_{a|x}] \\
{\rm s.t.} \qquad & \sum_\lambda \delta_{a,\lambda_x} \chi[G_\lambda]\ \succeq \chi[\Eax]\quad\forall a,x,\\
&\chi[G_\lambda]\succeq 0\quad\forall\lambda,\\
&\sum_a \chi[\Eax]=\chi[\openone]\quad\forall x,\\
&\chi[\Eax]\succeq 0\quad\forall a,x,\\
&\sum_\lambda \chi[G_\lambda]_{\openone} = \IR_0 + 1,\\
&I(P_0) = \sum_{a,x} \alpha_{a,x} \tr\left[\chi[\Eax]\ W_{a,x}\right],
\end{aligned}
\label{eq:feasib}
\end{equation}
where the matrices $W_{a,x}$ and the coefficients $\alpha_{a,x}$ are properly chosen to extract the Bell expression from the terms corresponding to probabilities appearing in $\{\chi[\Eax]\}_{a,x}$. It is then clear, then, the (nontrivial) extreme points of this set are equivalently characterized by the following two problems:
\begin{equation}\label{eq:minmax}
\begin{split}
&\text{ minimize } \IR \text{ given } I(P), \quad \text{ and }\\
&\text{ maximize } I(P) \text{ given } \IR.
 \end{split}
\end{equation} 
In fact, one may have highly incompatible observables and fail to obtain a highly violation of a Bell inequality due to the low entanglement in the shared state. The problems in Eq.~\eqref{eq:minmax} can be directly solved by transforming the feasibility problem in Eq.~\eqref{eq:feasib}. By construction, a feasible solution of one problem is also a feasible solution for the other one, so they characterize the same set of pairs $(\IR, I(P))$.
It is important to remark that here we are not using the full duality properties of the SDP, but simply the relation between $\IR$ and  $I(P)$ encoded in Eq.~\eqref{eq:feasib} and the fact that the problems in Eq.~\eqref{eq:minmax} are sufficient to characterize the nontrivial part of this set.

The formulation with the fixed $\IR_0$, however, provides an advantage since an extra condition can be imposed.  In fact, the substitution $\sum_\lambda \chi[G_\lambda]_{\openone} -1 = \IR_0$, allows us to write the third constraint of Eq.~\eqref{Eq_DIIR_app} as  $\sum_\lambda \chi[G_\lambda] = (\IR_0+1) \chi[\openone]$, effectively removing the nonlinearity appearing in the SDP in Eq.~\eqref{Eq_DIIR_app}. We then have
\begin{equation}
\begin{aligned}
\text{ Given } \qquad & \IR_0 \\
\max_{\chi[\openone],\{\chi[G_\lambda] , \chi[E_{a|x}] \}_{\lambda,a,x} } \quad & I(P)= \sum_{a,x} \alpha_{a,x} \tr\left[\chi[\Eax]\ W_{a,x}\right]\\
{\rm s.t.} \quad & \sum_\lambda \delta_{a,\lambda_x} \chi[G_\lambda]\ \succeq \chi[\Eax]\quad\forall a,x,\\
&\chi[G_\lambda]\succeq 0\quad\forall\lambda,\\
&\sum_\lambda \chi[G_\lambda] = (\IR_0+1)\chi[\openone],\\
&\sum_a \chi[\Eax]=\chi[\openone]\quad\forall x,\\
&\chi[\Eax]\succeq 0\quad\forall a,x,\\
&\sum_\lambda \chi[G_\lambda]_{\openone} = \IR_0 + 1,
\end{aligned}
\label{Eq_max_bell_app}
\end{equation}

The fact that the SDP in Eq.~\eqref{Eq_max_bell_app} provides a better characterization of the set $(\IR, I(P))$ is confirmed by numerical calculations. First, the incompatibility robustness has been analyzed in Fig.~\ref{Fig_comparison_DIIR_tCHSH}, where this distinction is not relevant. However, a characterization analogous to that in Eq.~\eqref{eq:minmax} appears also for the genuine multipartite incompatibility robustness. For that case, we can see directly that the use of the two different formulations provides different results and that the computation for a fixed robustness $\IR_0$ provides a better bound. More details can be found in App.~\ref{app:GMI}.


\section{SDP formulation for genuine-multipartite incompatibility}\label{app:GMI}


In the following, we recall several results from Ref.~\cite[]{QuintinoPRL2019}, in particular the SDPs \eqref{SDP_GMI} and \eqref{SDP_GMIR_app}, and discuss their DI relaxation via the MMM. 

Following \cite[]{QuintinoPRL2019}, we recall that genuine triplewise incompatibility , namely, the impossibility of writing
\begin{equation}
 E_{a|x}=p_{12} J^{12}_{a|x} + p_{23} J^{23}_{a|x} + p_{13} J^{13}_{a|x} 
\end{equation}
for some probabilities $p_{12}$, $p_{23}$, and $p_{13}$ with $p_{12}+p_{23}+p_{13}=1$, is equivalent to the infeasibility of the following SDP:
\begin{align}\label{SDP_GMI}
\text{Given}& \quad \{ E_{a|1} \}_a, \{E_{a|2}\}_a, \{E_{a|3}\}_a \\
\text{find}& \quad J^{12}_{a|x}\ , \; J^{23}_{a|x}\ , \; J^{31}_{a|x}\ , \; p_{12},\; p_{23},\; p_{31},\; G^{12}_{\lambda}, \; G^{23}_{\lambda}, \; G^{31}_{\lambda} \nonumber \\
\text{s.t.}&\quad G_\lambda^{12},G_\lambda^{23},G_\lambda^{13}\succeq 0, \; \; p_{12}, p_{23}, p_{13}	\geq 0, \nonumber\\ 
&\quad E_{a|x}=J^{12}_{a|x}+J^{23}_{a|x} + J^{13}_{a|x}\ ,\ \forall a,x, \nonumber\\ 
&\quad J^{12}_{a|x}\succeq 0, \; \forall a,x ; \; \sum_a J^{12}_{a|x} = p_{12} \openone, \; \forall x, \nonumber\\ 
&\quad J^{12}_{a|x} = \sum_\lambda \delta_{a,\lambda_x} G^{12}_\lambda \quad \text{ for } x=1,x=2 , \nonumber \\
&\quad J^{23}_{a|x}\succeq 0, \; \forall a,x ; \; \sum_a J^{23}_{a|x} = p_{23} \openone, \; \forall x, \; \nonumber\\ 
&\quad J^{23}_{a|x} = \sum_\lambda \delta_{a,\lambda_x} G^{23}_\lambda \quad \text{ for } x=2,x=3 ,\nonumber \\
&\quad J^{13}_{a|x}\succeq 0, \; \forall a,x ; \; \sum_a J^{13}_{a|x} = p_{13} \openone, \; \forall x, \nonumber\\ 
&\quad J^{13}_{a|x} = \sum_\lambda \delta_{a,\lambda_x} G^{13}_\lambda \quad \text{ for } x=1,x=3, \nonumber 
\end{align}
 where $\delta_{a,\lambda_x} $ is the deterministic strategy that assign probability $1$ if the $x$-th component of $\lambda$ is equal to $a$.

One can quantify the triplewise incompatibility of a set of measurements using SDP methods. We need few definitions and 
properties: $J^3_{a|x}:= J^{sx}_{a|x}+J^{tx}_{a|x} + J^{st}_{a|x}$ , for $x=1,2,3$ and $s,t, x$ all different. When 
$\{J_{a|x}^{12},J_{a|x}^{13},J_{a|x}^{23}\}_{a,x}$ is a solution of the problem in Eq.~\eqref{SDP_GMI}, we have that 
$J^{sx}_{a|x}$ and $J^{tx}_{a|x}$ arise each from a joint measurement, $J^{st}_{a|x}$ is positive, 
$\{G_\lambda^{st}\}$ is proportional to a POVM with the same proportionality constant as $\{J^{st}_{a|x}\}_a$ for
 all $x$, i.e., $\sum_{a} J^{st}_{a|x} = \sum_{\lambda} G^{st}_\lambda$ for all $s,t,x$. Finally, both 
 $\{J^3_{a|x}\}_a$ and $\{G^{12}_\lambda + G^{13}_\lambda + G^{23}_\lambda\}$ are POVMs.
From the above formulation, we can define a robustness with respect to arbitrary noise as 
\begin{equation}\label{eq:def_rob}
\begin{split}
t^*=\min \left\lbrace\ t \ \middle| J^3_{a|x} = \frac{E_{a|x} + t N_{a|x}}{1+t}, \text{ for } \{J^3_{a|x}\} \text{ sol.\ of} \text{ ~\eqref{SDP_GMI}, } \right.\\
\left. \{N_{a|x}\} \text{ meas. assemb. }\right\rbrace
\end{split}
\end{equation}
Following the argument in Ref.~\cite[]{QuintinoPRL2019}, one shows that $t*$ can be computed as
\begin{align}\label{SDP_GMIR_app}
 \text{Given }& \{E_{a|x}\}_{a,x}, \nonumber \\ 
 \text{and variables } &\{ G^{12}_\lambda , G^{13}_\lambda , G^{23}_\lambda \}_\lambda, \nonumber
 \{ J_{a|3}^{12},J_{a|2}^{13},J_{a|1}^{23} \}_{a} ,  \nonumber\\
\min\ & \frac{1}{d}\sum_\lambda \tr[G^{12}_\lambda + G^{13}_\lambda + G^{23}_\lambda ] - 1\\
\text{s.t. } & G^{st}_{\lambda} \succeq 0\ \forall \lambda,\ \sum_\lambda G^{st}_\lambda = \frac{\openone}{d} \sum_\lambda \tr[G^{st}_\lambda]\nonumber\\
& \text{ for } (s,t)=(1,2),(1,3),(2,3); \nonumber \\
& J^{st}_{a|x} \succeq 0, \ \forall a,\  \sum_a J^{st}_{a|x} = \sum_\lambda G^{st}_\lambda \text{ and }\nonumber \\
& \sum_\lambda \delta_{a,\lambda_x} (G^{sx}_\lambda + G^{tx}_\lambda) + J^{st}_{a|x} \succeq E_{a|x},\nonumber\\
& \text{ for } (s,t,x)=(1,2,3),(1,3,2),(2,3,1); \nonumber 
\end{align}	
to show the strict feasibility, implying via Slater's conditon that the primal and the dual problem have the same optimal values,  it is sufficient to take each $G^{st}_\lambda = \openone$ and the corresponding $J^{st}_{a|x}$ coming from the linear constraints.

Clearly, the same argument can be extended to define genuine multipartite incompatibility beyond the triplewise case.

\begin{figure}[t]
\emph{\includegraphics[width=0.45\textwidth]{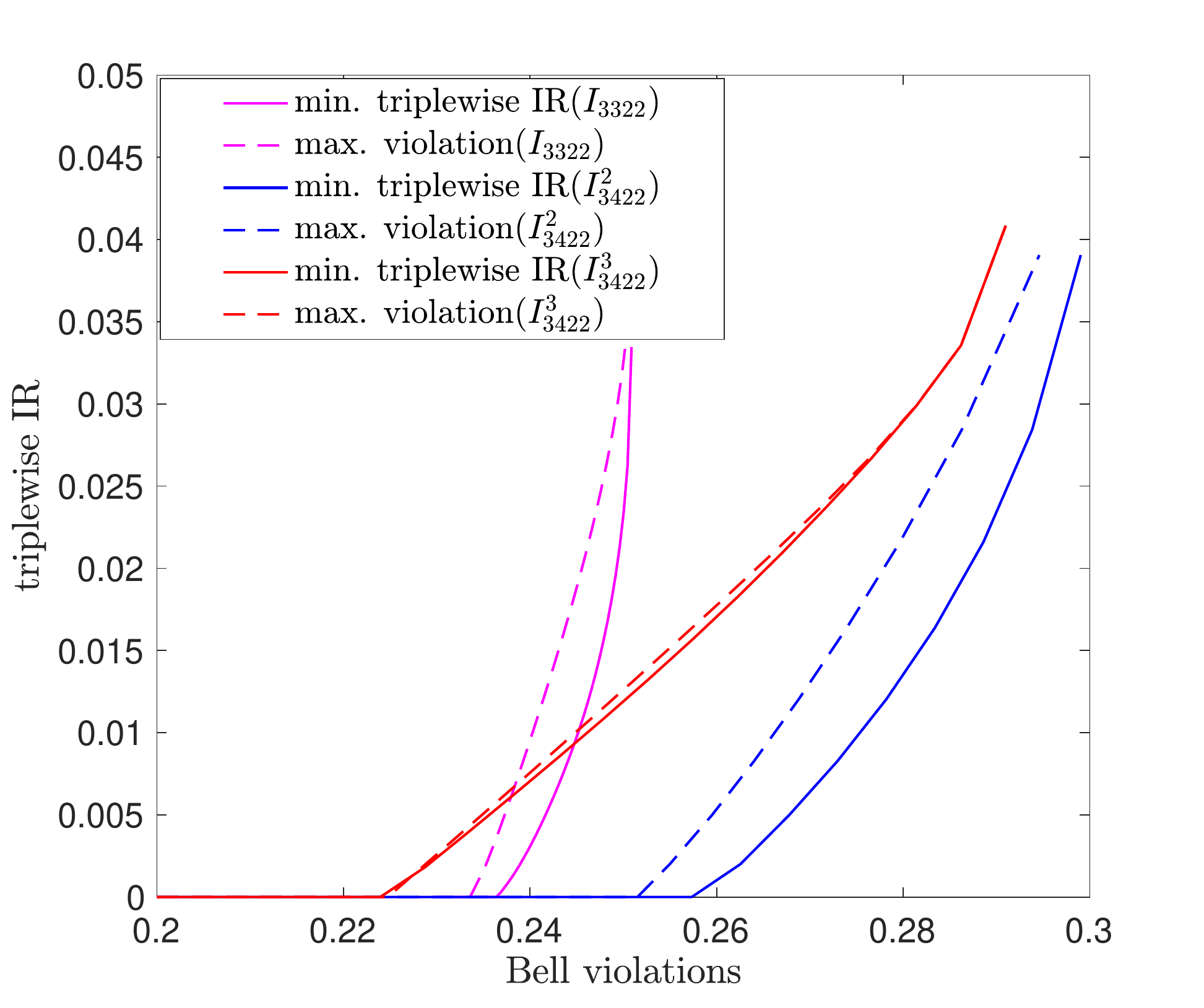} }
\caption{ Comparison between two SDPs computing the bounds. The solid curves are bounds obtained by fixing a Bell violation and minimizing the robustness (cf. Eq.~\eqref{SDP_DI_GMIR}) while the dashed curves are bounds obtained by fixing a robustness and maximizing the Bell violation (cf. Eq.~\eqref{SDP_DI_GMIR_fix_rob_app}).
}
\label{Fig_comparison_maxBell_and_minIR}
\end{figure}

Finally, we can show that the SDP computing the maximum of a Bell inequality $I(P)$ for a given robustness $\IR_0$, namely,
\begin{align}\label{SDP_DI_GMIR_fix_rob_app}
\text{Given:}\qquad & \IR_0,\nonumber \text{ and }\\
\text{variables:}\ & \{\chi[E_{a|x}]\}_{a,x}, \text{ and }\{\chi[G^{st}_\lambda]\}_\lambda, \{\chi[J^{st}_{a|x}]\}_{a},  \nonumber\\
&\text{ for } (s,t,x) = (1,2,3), (1,3,2),(2,3,1), \nonumber\\
\max\ & I(P) \nonumber \\
\text{s.t. } & \chi[G^{st}_{\lambda}] \succeq 0\ \forall \lambda, (s,t)=(1,2),(1,3),(2,3); \nonumber \\ 
&\sum_{\lambda,(s,t)} \chi[G^{st}_\lambda] =  \chi[\openone] (\IR_0+1)\ \text{ and } \nonumber\\
&\sum_{\lambda,(s,t)} \chi[G^{st}_\lambda]_{\openone} =  (\IR_0+1),\nonumber \\
& \text{ with sum over } (s,t)=(1,2),(1,3),(2,3); \nonumber \\
& \chi[J^{st}_{a|x}] \succeq 0, \ \forall a,\  \sum_a \chi[J^{st}_{a|x}] = \sum_\lambda \chi[G^{st}_\lambda] \text{ and }\nonumber \\
& \sum_\lambda \delta_{a,\lambda_x} (\chi[G^{sx}_\lambda] + \chi[G^{tx}_\lambda]) + \chi[J^{st}_{a|x}] \succeq \chi[E_{a|x}],\nonumber\\
& \text{ for } (s,t,x)=(1,2,3),(1,3,2),(2,3,1); \nonumber \\
& \chi[E_{a|x}] \succeq 0, \text{ for all } a,x, \nonumber \\
& \sum_a \chi[E_{a|x}] = \chi[\openone], \text{ for all } x,
\end{align}	
provides a better bound with respect to similar SDP computing the minimal robustness $\IR$ for a given Bell violation $I(P_0)$. More details can be found in Fig.~\ref{Fig_comparison_maxBell_and_minIR}. 

\section{Witnesses of genuine-multipartite incompatibility}\label{app:Wit_GMI}


To make our discuss self-contained, we briefly recall in this section two witnesses of GMI presented in Ref.~\cite[]{QuintinoPRL2019}. To keep the notation lighter, we will discuss only the case of genuine tripartite incompatibility; the argument can then be generalized to more measurements.

The authors of Ref.~\cite[]{QuintinoPRL2019} define the set $L_{12}^{Q}$ as the set of bipartite correlations with three measurements for Alice, in which the pair $x=1,2$ is compatible (one should specify also Bob's settings and outcomes, i.e., the whole Bell scenario). Given three measurement $\{\{E_{a|x}\}_a\}_{x=1,2,3}$ we say that\footnote{To simplify the notation, we use $P(ab|xy)$ to represent $\{P(a,b|x,y)\}_{a,b,x,y}$ when there is no risk of confusion.}
\begin{align}
& P(ab|xy) \text{ belongs to } L_{12}^Q \text{ if: } \nonumber \\
&P(ab|xy) \in \mathcal{Q}, \nonumber\\
&\{ \{P(ab|xy)\}_{a,b,y} \}_{x=1,2} \text{ is local }.
\end{align}
For the set of quantum correlations $\mathcal{Q}$, typically only an approximate characterization is possible, namely, via the NPA hierarchy of a given level $l$.

In particular, for the case of Alice having only dichotomic outcomes, the condition that $\{ \{P(ab|xy)\}_{a,b,y} \}_{x=1,2}$ is local can be simply imposed by requiring that all CHSH inequalities for Alice's pair of measurements and all possible pairs of dichotomized measurements for Bob are satisfied \cite[]{Pironio_2014}, namely
\begin{align}
& P(ab|xy) \text{ belongs to } L_{12}^Q \text{ if: } \nonumber \\
&P(ab|xy) \in \mathcal{Q}, \nonumber\\
&\{ \{P(ab|xy)\}_{a,b,y} \}_{x=1,2} \text{ satisfies CHSH for any }  \\
&\text{ pair of dichotomized measurements for Bob}.\nonumber
\end{align}

In simple terms, this set is obtained by the NPA-hierarchy constraints plus linear constraints corresponding to Bell inequalities involving only $x=1,2$ and all possible dichotomized measurements on Bob's side.

The above definition can be extended to the convex hull of $L_{12}, L_{13}, L_{23}$, i.e., $L_{2{\rm conv}}$ as follows
\begin{align}
& P(ab|xy) \text{ belongs to } L_{2{\rm conv}}^Q \text{ if: } \nonumber \\
&P(ab|xy) \in \mathcal{Q}, \nonumber\\
&P(ab|xy)= \mu_{12} P_{12}(ab|xy) + \mu_{13} P_{13}(ab|xy) + \mu_{23} P_{23}(ab|xy), \nonumber   \\
&P_{ij}(ab|xy) \in L^Q_{ij},\ \mu_{ij}\geq 0,\ \mu_{12}+\mu_{13}+\mu_{23}=1
\end{align}

As noticed in Ref.~\cite[]{QuintinoPRL2019}, imposing locality constraints at the level of the observed distribution is not the same as imposing constraints on the joint measurability of observables in the NPA hierarchy approximating the set $Q$. For instance, consider the set $Q_{12_{JM}}$ defined as follow
\begin{align}
& P(ab|xy) \text{ belongs to } Q_{12_{JM}} \text{ if: } \nonumber \\
&P(ab|xy) \in \mathcal{Q}, \text{ with } \nonumber\\
& E_{a|1} = \sum_{a'} M_{aa'}^{12},\ \forall a, \ E_{a'|2} = \sum_a M_{aa'}^{12}, \forall a'.\\
\end{align}

In other words, the two measurements $\{E_{a|1}\}_a$ and $\{E_{a|2}\}_a$ are substituted by a single joint measurement $M_{aa'}^{12}$. In terms of the NPA hierarchy, this can be simply obtained by taking the moments involving $M_{aa'}^{12}$ instead of $\{E_{a|1}\}_a$ and $\{E_{a|2}\}_a$. Similarly, the convex hull $Q_{2{\rm conv}_{JM}}$ can be defined as 
\begin{align}\label{eq:Q_conv}
& P(ab|xy) \text{ belongs to } Q_{2{\rm conv}_{JM}} \text{ if: } \nonumber \\
&P(ab|xy) \in \mathcal{Q}, \nonumber\\
&P(ab|xy)= \mu_{12} P_{12}(ab|xy) + \mu_{13} P_{13}(ab|xy) + \mu_{23} P_{23}(ab|xy), \nonumber   \\
&P_{ij}(ab|xy) \in Q_{ij_{JM}},\ \mu_{ij}\geq 0,\ \mu_{12}+\mu_{13}+\mu_{23}=1.
\end{align}
The SDP approximation of this set involves computing three different NPA moment matrices, one for each distribution $P_{ij}(ab|xy)$. 

It is important to remark that the NPA hierarchy can be computed by assuming the dilation of the POVMs to projective measurements. 
It is also important to remark that, even if some structure of measurement incompatibility require POVMs (e.g., the hollow triangle), in Eq.~\eqref{eq:Q_conv} only pairwise JM conditions arise, one for each $P_{ij}$. A total JM measurability condition among a measurement assemblage $\{ E_{a|x} \}_{a,x}$ is equivalent to the existence of a common dilation in which the measurements are represented by commuting projective measurements. In this sense, due to the convex nature of the genuine multipartite incompatibility problem, there is no contradiction between the use of the dilation and the fact that non-trivial compatibility structures necessarily require POVMs.


\section{Relation between the measurement moment matrix and the assemblage moment matrix}\label{app:previous_methods}


The \emph{assemblage moment matrices} proposed in Ref.~\cite[]{CBLC16} can be viewed as a special case of the MMM, as we show below. If the sequence $\{S_i\}$ in Eq.~(5) of the main text is only composed of Bob's projectors and their products, namely, $\{S_i\} = \{\openone\otimes B_i\}$ with $\{B_i\}=\{\openone, E_{1|1}^{\rm B}, E_{2|1}^{\rm B}, E_{1|1}^{\rm B}E_{1|2}^{\rm B}, {\rm etc}...\}$, then Eq.~(5) of the main text will be
\begin{equation}
\chi = \sum_{ij}\ket{i}\bra{j} \tr(B_j^\dag B_i \sigma_{a|x}),
\end{equation}
with $\sigma_{a|x}:=\tr_{\rm A}(\Eax\otimes\openone^{\rm B} \rab)$ being the \emph{state assemblage} in a 
steering-type experiment, which recovers the form of the assemblage moment matrices. Moreover, since each constraint in the SDP for computing the bounds in Ref.~\cite[]{CBLC16} is also a constraint of the SDP derived from Eq.~(7) in the main text, but not vice versa, the MMM bounds will never be worse than those in Ref.~\cite[]{CBLC16}. In Ref.~\cite[]{CBLC18}, the authors further obtained tighter DI bounds on $\IR$ by bounding another measure of steerability --- the \emph{consistent steering robustness}, which is also a lower bound on $\IR$~\cite[]{Cavalcanti16}. If we consider again that the sequence $\{S_i\}$ is only composed of Bob's part, the only difference between the SDP derived from Eq.~(7) in the main text and the SDP used for bounding the consistent steering robustness in Ref.~\cite[]{CBLC18} is that the latter does not include the fourth constraint of the former: $\sum_a \chi[\Eax]=\chi[\openone]$. As a consequence, the present DI bound on $\IR$ will not be lower than that of Ref.~\cite[]{CBLC18}.

\begin{figure}[t]
\includegraphics[width=1\linewidth]{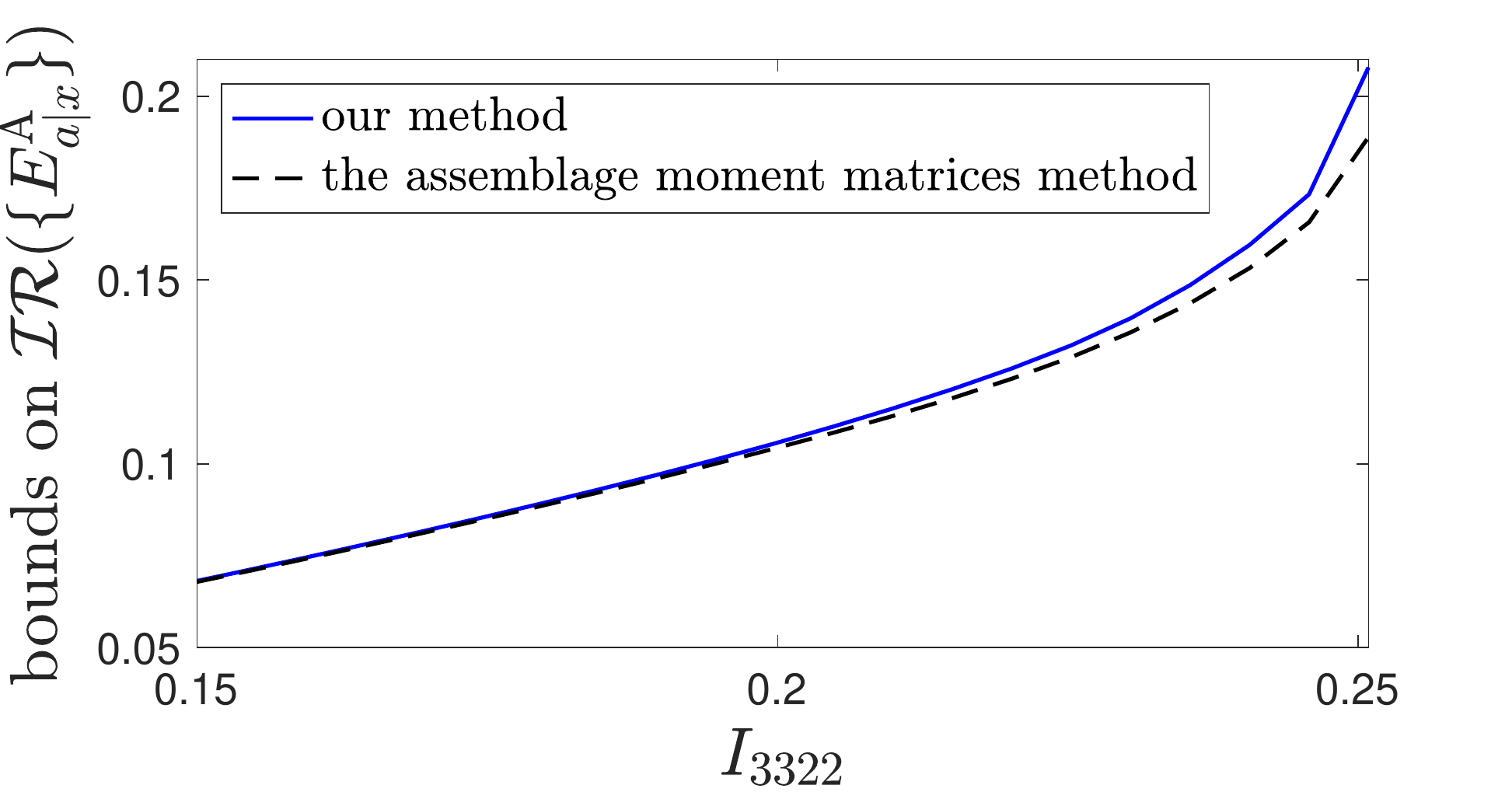}
\caption{Comparison between lower bounds on $\IR$ in the $I_{3322}$ scenario~\cite[]{Collins04}. The blue-solid and black-dashed curves represent, respectively, lower bounds obtained from our method and from the method of the \emph{assemblage moment matrices}~\cite[]{CBLC16}. The local and quantum bounds for the $I_{3322}$ inequality are, respectively, $0$ and around $0.250875561$~\cite[]{Rosset18SymDPoly}. The levels of the hierarchy of the semidefinite relaxation used to carry out the computation in both methods are the $3$rd level.
}
\label{Fig_comparison_DIIR_I3322_level3}
\end{figure}

Finally, by computing explicitly the $\IR$ bounds associated with a given violation of the $I_{3322}$ inequality and provided, respectively, by the method in Ref.~\cite{CBLC16} and by the MMM method, we show that the MMM method provides a tighter value. The results of numerical calculations are plotted in Fig.~\ref{Fig_comparison_DIIR_I3322_level3}.


\section{POVMs and projective measurements in the SDI scenario}\label{app:POVM_dil}
Projective measurements, via their idempotence and orthogonality properties ($P_a P_b = P_a \delta_{a,b}$), allow for a great simplification of the sequences appearing in the construction of moment matrices. In the DI scenario, all measurements can be assumed to be projective due to the Neumark dilation, as discussed in the main text. Such a dilation, however, requires to increase the Hilbert space dimension and is, thus, not always possible if the dimension of the system is constrained as in the SDI scenario. In some cases, however, projective measurements can be recovered by a convexity argument. For instance, for dichotomic measurements, it is known that they are all convex mixtures of projective measurements (intuitively, it is sufficient to decompose the $0$-outcome element), so we can restrict ourselves to projective measurements if the objective function we wish to minimize is linear in the POVMs operator. This is the case for, e.g., Bell inequalities as noted in \cite[]{Navascues15prl}, but it is also the case for the incompatibility robustness. In order to show that, it is useful to introduce first some slack variables ($\{S_{a,x}\}_{a,x}$), namely,
\begin{equation}
\begin{aligned}
\IR +1 = \min_{\{G_\lambda\}} \quad & \frac{1}{d}\sum_\lambda \tr[G_\lambda]\\
{\rm s.t.}\quad  &  \ \sum_\lambda \delta_{a,\lambda_x} G_\lambda + S_{a,x} =\Eax\ \forall\ a,x,\ \\
& G_\lambda\succeq 0, \ S_{a,x} \succeq 0,\ \forall\ \lambda, a, x,\ \\
&\sum_\lambda G_\lambda -  \frac{1}{d}\left(\sum_\lambda\tr[G_\lambda]\right)\cdot \openone=0,
\end{aligned}
\end{equation}
to put the problem in the standard form
\begin{equation}
\begin{aligned}
\min_{X} \quad & \mean{C,X}\\
{\rm s.t.} \quad &  \mean{A_k, X}=b_k, \forall\ k\\
& X \succeq 0.
\end{aligned}
\end{equation}
It is then clear that the entries of the POVM elements $E_{a|x}$ will appear in the vector $b$, and consequently in the objective of the dual problem
\begin{equation}
\begin{aligned}
\max_{y} \quad & \mean{b,y}\\
{\rm s.t.} \quad &  \sum_k y_k A_k\preceq C.
\end{aligned}
\end{equation}
It is clear that if, for a given $x$, $E_{a|x} = \sum_i \mu_i P_{a|x}^{i}$, the minimal robustness will be obtained for a given projective measurement $\{P^i_{a|x}\}_{a,x}$.

It is not obvious, however, what happens if one tries to minimize the robustness for a fixed Bell inequality violation. In fact, by choosing one element of the decomposition as above, we may decrease both the robustness and the Bell inequality violation. 

A possible approach to the problem by dilation of the measurements of both Alice and Bob, has been already proposed in Ref.~\cite[]{Navascues15prl}, while generating a basis for $\mathcal{S}_d$, the space of moment matrices corresponding to dimension $d$, one should sample Alice's and Bob's measurements of the form
\beq\label{eq:dilation_sdi}
&&\Eax = U^x(\ket{a}\bra{a}_{A'}\otimes \openone_{A})(U^x)^\dagger,\nonumber\\
&&\Eby = U^y(\ket{b}\bra{b}_{B'}\otimes \openone_{B})(U^y)^\dagger,
\eeq
with random unitaries $U^x$ and $U^y$. Random states should then be taken of the form $\rho = \ket{0}\bra{0}_{A'}\otimes\ket{0}\bra{0}_{B'}\otimes\ket{\psi}\bra{\psi}_{AB}$. Since we are interested in dichotomic measurements, dimension of the auxiliary spaces $A'$ and $B'$ is $2$ in both cases. 
In practice, however, this method was not able to provide a better bound of the SDI bound in $d=2$ for the $I_{3322}$ inequality.

\section{Extension of MMM method to the prepare-and-measure scenario}\label{app:SDI}
The prepare-and-measure (P-M) scenario, e.g., the one given by random access codes~\cite[]{PawlowskiPRA2011,Carmeli_2020}, is a paradigm often considered in quantum information processing as an alternative to the Bell scenario. The P-M scenario is a one-way communication scenario in which one party, let us say Bob, prepares a physical system in a state $\rho_y$ chosen from a finite set indexed by $y$ and sends it to the other party, Alice. Alice measures this system with a choice of measurement specified by $x$. The conditional distribution $P(a|x,y)$, where $a$ is the outcome of Alice's measurement, is then used to semi-device-independently characterize the states and measurements in this scenario. The classical distribution $P(a|x,y)$ is the one produced by states and measurements which can be simultaneously diagonalized in some basis of the Hilbert space in which they are defined. 

 One important distinction between the P-M and Bell scenario is that parties' measurements do not need to be space-like separated. However, in order to observe a gap between classical and quantum strategies some form of restriction on the communication needs to be imposed~\cite[]{Hoffmann_2018,Budroni_2019}. Here, we consider the most common type of restriction, an upper-bound on the Hilbert's space dimension in which the states and measurements are defined. This enables us to use the hierarchy of Ref.~\cite[]{Navascues15prl} to approximate the set of quantum correlations $P(a|x,y)$ and subsequently map the incompatibility robustness SDP to MMM SDP.
 
 The map is a direct extension, merely a simplification of Eq.~(4) of the main text, and can be written as follows:
\begin{equation}
\chi[\Eax]:=\sum_n K_n (\Eax) K_n^\dag\quad\forall a,x,
\label{Eq_CP_map_SDI}
\end{equation}
where $K_n:=\sum_i \ket{i}_{\overline{\rm A}{\rm A}}\bra{n}S_i$, and $\{S_i\}$ is the following sequence of operators: 
$\{S_i\} = \{\openone^{\rm A},\Eax,\rho_y, \Eax\rho_y,  {\rm etc.} \}$. The MMM can then be defined as
\begin{equation}
\chi_{\{S_i\}}[\Eax] = \sum_{ij}\ket{i}\bra{j}\tr\left[S_i(\Eax)S_j^\dag\right],
\label{Eq_LM_SDI}
\end{equation}
which is a direct analogy of Eq.~(5) of the main text. Using this map, one can formulate an SDI relaxation of incompatibility robustness SDP, which reads
\begin{equation}
\begin{aligned}
\min_{\{\chi[G_\lambda] , \chi[E_{a|x}] \}_{\lambda,a,x} } \quad & \sum_\lambda \chi[G_\lambda]_{\openone} -1\\
{\rm s.t.} \quad & \sum_\lambda \delta_{a,\lambda_x} \chi[G_\lambda]\ \succeq \chi[\Eax]\quad\forall a,x,\\
&\chi[G_\lambda]\succeq 0\quad\forall\lambda,\\
&\sum_\lambda \chi[G_\lambda] = \sum_\lambda \chi[G_\lambda]_{\openone}\cdot\chi[\openone],\\
&\sum_a \chi[\Eax]=\chi[\openone]\quad\forall x,\\
&\chi[\Eax]\succeq 0\quad\forall a,x,\\
&\chi[\Eax]\in \mathcal{S}_d, \quad\forall a,x,\\
&\chi[\openone]\in \mathcal{S}_d,\\
&P(a|x,y) = P_{\rm obs}(a|x,y)\quad\forall a,x,y,
\end{aligned}
\label{Eq_DIIR_SDI}
\end{equation}
where $\mathcal{S}_d$ is a subspace of moment matrices spanned by those corresponding to states and measurements defined on Hilbert space of dimension $d$.


%



\begin{thebibliography}{70}%
\makeatletter
\providecommand \@ifxundefined [1]{%
 \@ifx{#1\undefined}
}%
\providecommand \@ifnum [1]{%
 \ifnum #1\expandafter \@firstoftwo
 \else \expandafter \@secondoftwo
 \fi
}%
\providecommand \@ifx [1]{%
 \ifx #1\expandafter \@firstoftwo
 \else \expandafter \@secondoftwo
 \fi
}%
\providecommand \natexlab [1]{#1}%
\providecommand \enquote  [1]{``#1''}%
\providecommand \bibnamefont  [1]{#1}%
\providecommand \bibfnamefont [1]{#1}%
\providecommand \citenamefont [1]{#1}%
\providecommand \href@noop [0]{\@secondoftwo}%
\providecommand \href [0]{\begingroup \@sanitize@url \@href}%
\providecommand \@href[1]{\@@startlink{#1}\@@href}%
\providecommand \@@href[1]{\endgroup#1\@@endlink}%
\providecommand \@sanitize@url [0]{\catcode `\\12\catcode `\$12\catcode
  `\&12\catcode `\#12\catcode `\^12\catcode `\_12\catcode `\%12\relax}%
\providecommand \@@startlink[1]{}%
\providecommand \@@endlink[0]{}%
\providecommand \url  [0]{\begingroup\@sanitize@url \@url }%
\providecommand \@url [1]{\endgroup\@href {#1}{\urlprefix }}%
\providecommand \urlprefix  [0]{URL }%
\providecommand \Eprint [0]{\href }%
\providecommand \doibase [0]{http://dx.doi.org/}%
\providecommand \selectlanguage [0]{\@gobble}%
\providecommand \bibinfo  [0]{\@secondoftwo}%
\providecommand \bibfield  [0]{\@secondoftwo}%
\providecommand \translation [1]{[#1]}%
\providecommand \BibitemOpen [0]{}%
\providecommand \bibitemStop [0]{}%
\providecommand \bibitemNoStop [0]{.\EOS\space}%
\providecommand \EOS [0]{\spacefactor3000\relax}%
\providecommand \BibitemShut  [1]{\csname bibitem#1\endcsname}%
\let\auto@bib@innerbib\@empty
\bibitem [{\citenamefont {Heisenberg}(1927)}]{Heisenberg1927}%
  \BibitemOpen
  \bibfield  {author} {\bibinfo {author} {\bibfnamefont {W.}~\bibnamefont
  {Heisenberg}},\ }\bibfield  {title} {\enquote {\bibinfo {title} {{\"U}ber den
  anschaulichen inhalt der quantentheoretischen kinematik und mechanik},}\
  }\href {\doibase 10.1007/BF01397280} {\bibfield  {journal} {\bibinfo
  {journal} {Z. Physik}\ ,\ \bibinfo {pages} {172–--198}} (\bibinfo {year}
  {1927})}\BibitemShut {NoStop}%
\bibitem [{\citenamefont {Robertson}(1929)}]{Robertson29}%
  \BibitemOpen
  \bibfield  {author} {\bibinfo {author} {\bibfnamefont {H.~P.}\ \bibnamefont
  {Robertson}},\ }\bibfield  {title} {\enquote {\bibinfo {title} {The
  uncertainty principle},}\ }\href {\doibase 10.1103/PhysRev.34.163} {\bibfield
   {journal} {\bibinfo  {journal} {Phys. Rev.}\ }\textbf {\bibinfo {volume}
  {34}},\ \bibinfo {pages} {163--164} (\bibinfo {year} {1929})}\BibitemShut
  {NoStop}%
\bibitem [{\citenamefont {Busch}\ \emph {et~al.}(2014)\citenamefont {Busch},
  \citenamefont {Lahti},\ and\ \citenamefont {Werner}}]{BLWRev}%
  \BibitemOpen
  \bibfield  {author} {\bibinfo {author} {\bibfnamefont {P.}~\bibnamefont
  {Busch}}, \bibinfo {author} {\bibfnamefont {P.}~\bibnamefont {Lahti}}, \ and\
  \bibinfo {author} {\bibfnamefont {R.~F.}\ \bibnamefont {Werner}},\ }\bibfield
   {title} {\enquote {\bibinfo {title} {\textit{Colloquium} : Quantum
  root-mean-square error and measurement uncertainty relations},}\ }\href
  {\doibase 10.1103/RevModPhys.86.1261} {\bibfield  {journal} {\bibinfo
  {journal} {Rev. Mod. Phys.}\ }\textbf {\bibinfo {volume} {86}},\ \bibinfo
  {pages} {1261--1281} (\bibinfo {year} {2014})}\BibitemShut {NoStop}%
\bibitem [{\citenamefont {Bell}(1964)}]{Bell64}%
  \BibitemOpen
  \bibfield  {author} {\bibinfo {author} {\bibfnamefont {J.~S.}\ \bibnamefont
  {Bell}},\ }\bibfield  {title} {\enquote {\bibinfo {title} {On the {E}instein
  {P}odolsky {R}osen paradox},}\ }\href {\doibase
  10.1103/PhysicsPhysiqueFizika.1.195} {\bibfield  {journal} {\bibinfo
  {journal} {Physics Physique Fizika}\ }\textbf {\bibinfo {volume} {1}},\
  \bibinfo {pages} {195--200} (\bibinfo {year} {1964})}\BibitemShut {NoStop}%
\bibitem [{\citenamefont {Brunner}\ \emph {et~al.}(2014)\citenamefont
  {Brunner}, \citenamefont {Cavalcanti}, \citenamefont {Pironio}, \citenamefont
  {Scarani},\ and\ \citenamefont {Wehner}}]{Brunner14}%
  \BibitemOpen
  \bibfield  {author} {\bibinfo {author} {\bibfnamefont {N.}~\bibnamefont
  {Brunner}}, \bibinfo {author} {\bibfnamefont {D.}~\bibnamefont {Cavalcanti}},
  \bibinfo {author} {\bibfnamefont {S.}~\bibnamefont {Pironio}}, \bibinfo
  {author} {\bibfnamefont {V.}~\bibnamefont {Scarani}}, \ and\ \bibinfo
  {author} {\bibfnamefont {S.}~\bibnamefont {Wehner}},\ }\bibfield  {title}
  {\enquote {\bibinfo {title} {Bell nonlocality},}\ }\href {\doibase
  10.1103/RevModPhys.86.419} {\bibfield  {journal} {\bibinfo  {journal} {Rev.
  Mod. Phys.}\ }\textbf {\bibinfo {volume} {86}},\ \bibinfo {pages} {419--478}
  (\bibinfo {year} {2014})}\BibitemShut {NoStop}%
\bibitem [{\citenamefont {Schr\"odinger}(1935)}]{Schrodinger35}%
  \BibitemOpen
  \bibfield  {author} {\bibinfo {author} {\bibfnamefont {E.}~\bibnamefont
  {Schr\"odinger}},\ }\bibfield  {title} {\enquote {\bibinfo {title}
  {Discussion of probability relations between separated systems},}\ }\href
  {http://journals.cambridge.org/article_S0305004100013554} {\bibfield
  {journal} {\bibinfo  {journal} {Proc. Cambridge Phil. Soc.}\ }\textbf
  {\bibinfo {volume} {31}},\ \bibinfo {pages} {555} (\bibinfo {year}
  {1935})}\BibitemShut {NoStop}%
\bibitem [{\citenamefont {Cavalcanti}\ and\ \citenamefont
  {Skrzypczyk}(2017)}]{Cavalcanti17}%
  \BibitemOpen
  \bibfield  {author} {\bibinfo {author} {\bibfnamefont {D.}~\bibnamefont
  {Cavalcanti}}\ and\ \bibinfo {author} {\bibfnamefont {P.}~\bibnamefont
  {Skrzypczyk}},\ }\bibfield  {title} {\enquote {\bibinfo {title} {Quantum
  steering: a review with focus on semidefinite programming},}\ }\href
  {http://stacks.iop.org/0034-4885/80/i=2/a=024001} {\bibfield  {journal}
  {\bibinfo  {journal} {Reports on Progress in Physics}\ }\textbf {\bibinfo
  {volume} {80}},\ \bibinfo {pages} {024001} (\bibinfo {year}
  {2017})}\BibitemShut {NoStop}%
\bibitem [{\citenamefont {Uola}\ \emph {et~al.}(2020)\citenamefont {Uola},
  \citenamefont {Costa}, \citenamefont {Nguyen},\ and\ \citenamefont
  {G\"uhne}}]{Uola2020Steering}%
  \BibitemOpen
  \bibfield  {author} {\bibinfo {author} {\bibfnamefont {R.}~\bibnamefont
  {Uola}}, \bibinfo {author} {\bibfnamefont {A.~C.~S.}\ \bibnamefont {Costa}},
  \bibinfo {author} {\bibfnamefont {H.~C.}\ \bibnamefont {Nguyen}}, \ and\
  \bibinfo {author} {\bibfnamefont {O.}~\bibnamefont {G\"uhne}},\ }\bibfield
  {title} {\enquote {\bibinfo {title} {Quantum steering},}\ }\href {\doibase
  10.1103/RevModPhys.92.015001} {\bibfield  {journal} {\bibinfo  {journal}
  {Rev. Mod. Phys.}\ }\textbf {\bibinfo {volume} {92}},\ \bibinfo {pages}
  {015001} (\bibinfo {year} {2020})}\BibitemShut {NoStop}%
\bibitem [{\citenamefont {Kochen}\ and\ \citenamefont {Specker}(1967)}]{KS67}%
  \BibitemOpen
  \bibfield  {author} {\bibinfo {author} {\bibfnamefont {S.}~\bibnamefont
  {Kochen}}\ and\ \bibinfo {author} {\bibfnamefont {E.}~\bibnamefont
  {Specker}},\ }\bibfield  {title} {\enquote {\bibinfo {title} {The problem of
  hidden variables in quantum mechanics},}\ }\href@noop {} {\bibfield
  {journal} {\bibinfo  {journal} {J. Math. Mech.}\ }\textbf {\bibinfo {volume}
  {17}},\ \bibinfo {pages} {59} (\bibinfo {year} {1967})}\BibitemShut {NoStop}%
\bibitem [{\citenamefont {Klyachko}\ \emph {et~al.}(2008)\citenamefont
  {Klyachko}, \citenamefont {Can}, \citenamefont {Binicio\u{g}lu},\ and\
  \citenamefont {Shumovsky}}]{KlyachkoPRL2008}%
  \BibitemOpen
  \bibfield  {author} {\bibinfo {author} {\bibfnamefont {A.~A.}\ \bibnamefont
  {Klyachko}}, \bibinfo {author} {\bibfnamefont {M.~A.}\ \bibnamefont {Can}},
  \bibinfo {author} {\bibfnamefont {S.}~\bibnamefont {Binicio\u{g}lu}}, \ and\
  \bibinfo {author} {\bibfnamefont {A.~S.}\ \bibnamefont {Shumovsky}},\
  }\bibfield  {title} {\enquote {\bibinfo {title} {Simple test for hidden
  variables in spin-1 systems},}\ }\href {\doibase
  10.1103/PhysRevLett.101.020403} {\bibfield  {journal} {\bibinfo  {journal}
  {Phys. Rev. Lett.}\ }\textbf {\bibinfo {volume} {101}},\ \bibinfo {pages}
  {020403} (\bibinfo {year} {2008})}\BibitemShut {NoStop}%
\bibitem [{\citenamefont {Cabello}(2008)}]{CabelloPRL2008}%
  \BibitemOpen
  \bibfield  {author} {\bibinfo {author} {\bibfnamefont {A.}~\bibnamefont
  {Cabello}},\ }\bibfield  {title} {\enquote {\bibinfo {title} {Experimentally
  testable state-independent quantum contextuality},}\ }\href {\doibase
  10.1103/PhysRevLett.101.210401} {\bibfield  {journal} {\bibinfo  {journal}
  {Phys. Rev. Lett.}\ }\textbf {\bibinfo {volume} {101}},\ \bibinfo {pages}
  {210401} (\bibinfo {year} {2008})}\BibitemShut {NoStop}%
\bibitem [{\citenamefont {Liang}\ \emph
  {et~al.}(2011{\natexlab{a}})\citenamefont {Liang}, \citenamefont {Spekkens},\
  and\ \citenamefont {Wiseman}}]{LIANG2011}%
  \BibitemOpen
  \bibfield  {author} {\bibinfo {author} {\bibfnamefont {Y.-C.}\ \bibnamefont
  {Liang}}, \bibinfo {author} {\bibfnamefont {R.~W.}\ \bibnamefont {Spekkens}},
  \ and\ \bibinfo {author} {\bibfnamefont {H.~M.}\ \bibnamefont {Wiseman}},\
  }\bibfield  {title} {\enquote {\bibinfo {title} {Specker’s parable of the
  overprotective seer: A road to contextuality, nonlocality and
  complementarity},}\ }\href {\doibase
  https://doi.org/10.1016/j.physrep.2011.05.001} {\bibfield  {journal}
  {\bibinfo  {journal} {Physics Reports}\ }\textbf {\bibinfo {volume} {506}},\
  \bibinfo {pages} {1 -- 39} (\bibinfo {year}
  {2011}{\natexlab{a}})}\BibitemShut {NoStop}%
\bibitem [{\citenamefont {{Budroni}}\ \emph {et~al.}(2021)\citenamefont
  {{Budroni}}, \citenamefont {{Cabello}}, \citenamefont {{G{\"u}hne}},
  \citenamefont {{Kleinmann}},\ and\ \citenamefont
  {{Larsson}}}]{Context_review}%
  \BibitemOpen
  \bibfield  {author} {\bibinfo {author} {\bibfnamefont {C.}~\bibnamefont
  {{Budroni}}}, \bibinfo {author} {\bibfnamefont {A.}~\bibnamefont
  {{Cabello}}}, \bibinfo {author} {\bibfnamefont {O.}~\bibnamefont
  {{G{\"u}hne}}}, \bibinfo {author} {\bibfnamefont {M.}~\bibnamefont
  {{Kleinmann}}}, \ and\ \bibinfo {author} {\bibfnamefont {J.-{\r{A}}.}\
  \bibnamefont {{Larsson}}},\ }\bibfield  {title} {\enquote {\bibinfo {title}
  {{Quantum Contextuality}},}\ }\href@noop {} {\bibfield  {journal} {\bibinfo
  {journal} {arXiv}\ } (\bibinfo {year} {2021})},\ \Eprint
  {http://arxiv.org/abs/2102.13036} {arXiv:2102.13036 [quant-ph]} \BibitemShut
  {NoStop}%
\bibitem [{\citenamefont {Wolf}\ \emph {et~al.}(2009)\citenamefont {Wolf},
  \citenamefont {Perez-Garcia},\ and\ \citenamefont {Fernandez}}]{Wolf09}%
  \BibitemOpen
  \bibfield  {author} {\bibinfo {author} {\bibfnamefont {M.~M.}\ \bibnamefont
  {Wolf}}, \bibinfo {author} {\bibfnamefont {D.}~\bibnamefont {Perez-Garcia}},
  \ and\ \bibinfo {author} {\bibfnamefont {C.}~\bibnamefont {Fernandez}},\
  }\bibfield  {title} {\enquote {\bibinfo {title} {Measurements incompatible in
  quantum theory cannot be measured jointly in any other no-signaling
  theory},}\ }\href {\doibase 10.1103/PhysRevLett.103.230402} {\bibfield
  {journal} {\bibinfo  {journal} {Phys. Rev. Lett.}\ }\textbf {\bibinfo
  {volume} {103}},\ \bibinfo {pages} {230402} (\bibinfo {year}
  {2009})}\BibitemShut {NoStop}%
\bibitem [{\citenamefont {Quintino}\ \emph {et~al.}(2014)\citenamefont
  {Quintino}, \citenamefont {V\'ertesi},\ and\ \citenamefont
  {Brunner}}]{Quint14}%
  \BibitemOpen
  \bibfield  {author} {\bibinfo {author} {\bibfnamefont {M.~T.}\ \bibnamefont
  {Quintino}}, \bibinfo {author} {\bibfnamefont {T.}~\bibnamefont {V\'ertesi}},
  \ and\ \bibinfo {author} {\bibfnamefont {N.}~\bibnamefont {Brunner}},\
  }\bibfield  {title} {\enquote {\bibinfo {title} {Joint measurability,
  {E}instein-{P}odolsky-{R}osen steering, and {B}ell nonlocality},}\ }\href
  {\doibase 10.1103/PhysRevLett.113.160402} {\bibfield  {journal} {\bibinfo
  {journal} {Phys. Rev. Lett.}\ }\textbf {\bibinfo {volume} {113}},\ \bibinfo
  {pages} {160402} (\bibinfo {year} {2014})}\BibitemShut {NoStop}%
\bibitem [{\citenamefont {Uola}\ \emph {et~al.}(2014)\citenamefont {Uola},
  \citenamefont {Moroder},\ and\ \citenamefont {G\"uhne}}]{Uola14}%
  \BibitemOpen
  \bibfield  {author} {\bibinfo {author} {\bibfnamefont {R.}~\bibnamefont
  {Uola}}, \bibinfo {author} {\bibfnamefont {T.}~\bibnamefont {Moroder}}, \
  and\ \bibinfo {author} {\bibfnamefont {O.}~\bibnamefont {G\"uhne}},\
  }\bibfield  {title} {\enquote {\bibinfo {title} {Joint measurability of
  generalized measurements implies classicality},}\ }\href {\doibase
  10.1103/PhysRevLett.113.160403} {\bibfield  {journal} {\bibinfo  {journal}
  {Phys. Rev. Lett.}\ }\textbf {\bibinfo {volume} {113}},\ \bibinfo {pages}
  {160403} (\bibinfo {year} {2014})}\BibitemShut {NoStop}%
\bibitem [{\citenamefont {Xu}\ and\ \citenamefont {Cabello}(2019)}]{Xu19}%
  \BibitemOpen
  \bibfield  {author} {\bibinfo {author} {\bibfnamefont {Z.-P.}\ \bibnamefont
  {Xu}}\ and\ \bibinfo {author} {\bibfnamefont {A.}~\bibnamefont {Cabello}},\
  }\bibfield  {title} {\enquote {\bibinfo {title} {Necessary and sufficient
  condition for contextuality from incompatibility},}\ }\href {\doibase
  10.1103/PhysRevA.99.020103} {\bibfield  {journal} {\bibinfo  {journal} {Phys.
  Rev. A}\ }\textbf {\bibinfo {volume} {99}},\ \bibinfo {pages} {020103}
  (\bibinfo {year} {2019})}\BibitemShut {NoStop}%
\bibitem [{\citenamefont {Tavakoli}\ and\ \citenamefont
  {Uola}(2020)}]{Tavakoli2020Measurement}%
  \BibitemOpen
  \bibfield  {author} {\bibinfo {author} {\bibfnamefont {A.}~\bibnamefont
  {Tavakoli}}\ and\ \bibinfo {author} {\bibfnamefont {R.}~\bibnamefont
  {Uola}},\ }\bibfield  {title} {\enquote {\bibinfo {title} {Measurement
  incompatibility and steering are necessary and sufficient for operational
  contextuality},}\ }\href {\doibase 10.1103/PhysRevResearch.2.013011}
  {\bibfield  {journal} {\bibinfo  {journal} {Phys. Rev. Research}\ }\textbf
  {\bibinfo {volume} {2}},\ \bibinfo {pages} {013011} (\bibinfo {year}
  {2020})}\BibitemShut {NoStop}%
\bibitem [{\citenamefont {Carmeli}\ \emph {et~al.}(2019)\citenamefont
  {Carmeli}, \citenamefont {Heinosaari},\ and\ \citenamefont
  {Toigo}}]{Carmeli19}%
  \BibitemOpen
  \bibfield  {author} {\bibinfo {author} {\bibfnamefont {C.}~\bibnamefont
  {Carmeli}}, \bibinfo {author} {\bibfnamefont {T.}~\bibnamefont {Heinosaari}},
  \ and\ \bibinfo {author} {\bibfnamefont {A.}~\bibnamefont {Toigo}},\
  }\bibfield  {title} {\enquote {\bibinfo {title} {Quantum incompatibility
  witnesses},}\ }\href {\doibase 10.1103/PhysRevLett.122.130402} {\bibfield
  {journal} {\bibinfo  {journal} {Phys. Rev. Lett.}\ }\textbf {\bibinfo
  {volume} {122}},\ \bibinfo {pages} {130402} (\bibinfo {year}
  {2019})}\BibitemShut {NoStop}%
\bibitem [{\citenamefont {Skrzypczyk}\ \emph {et~al.}(2019)\citenamefont
  {Skrzypczyk}, \citenamefont {\ifmmode \check{S}\else
  \v{S}\fi{}upi\ifmmode~\acute{c}\else \'{c}\fi{}},\ and\ \citenamefont
  {Cavalcanti}}]{Skrzypczyk19}%
  \BibitemOpen
  \bibfield  {author} {\bibinfo {author} {\bibfnamefont {P.}~\bibnamefont
  {Skrzypczyk}}, \bibinfo {author} {\bibfnamefont {I.}~\bibnamefont {\ifmmode
  \check{S}\else \v{S}\fi{}upi\ifmmode~\acute{c}\else \'{c}\fi{}}}, \ and\
  \bibinfo {author} {\bibfnamefont {D.}~\bibnamefont {Cavalcanti}},\ }\bibfield
   {title} {\enquote {\bibinfo {title} {All sets of incompatible measurements
  give an advantage in quantum state discrimination},}\ }\href {\doibase
  10.1103/PhysRevLett.122.130403} {\bibfield  {journal} {\bibinfo  {journal}
  {Phys. Rev. Lett.}\ }\textbf {\bibinfo {volume} {122}},\ \bibinfo {pages}
  {130403} (\bibinfo {year} {2019})}\BibitemShut {NoStop}%
\bibitem [{\citenamefont {Uola}\ \emph {et~al.}(2019)\citenamefont {Uola},
  \citenamefont {Kraft}, \citenamefont {Shang}, \citenamefont {Yu},\ and\
  \citenamefont {G\"uhne}}]{Uola19a}%
  \BibitemOpen
  \bibfield  {author} {\bibinfo {author} {\bibfnamefont {R.}~\bibnamefont
  {Uola}}, \bibinfo {author} {\bibfnamefont {T.}~\bibnamefont {Kraft}},
  \bibinfo {author} {\bibfnamefont {J.}~\bibnamefont {Shang}}, \bibinfo
  {author} {\bibfnamefont {X.-D.}\ \bibnamefont {Yu}}, \ and\ \bibinfo {author}
  {\bibfnamefont {O.}~\bibnamefont {G\"uhne}},\ }\bibfield  {title} {\enquote
  {\bibinfo {title} {Quantifying quantum resources with conic programming},}\
  }\href {\doibase 10.1103/PhysRevLett.122.130404} {\bibfield  {journal}
  {\bibinfo  {journal} {Phys. Rev. Lett.}\ }\textbf {\bibinfo {volume} {122}},\
  \bibinfo {pages} {130404} (\bibinfo {year} {2019})}\BibitemShut {NoStop}%
\bibitem [{\citenamefont {Takagi}\ \emph {et~al.}(2019)\citenamefont {Takagi},
  \citenamefont {Regula}, \citenamefont {Bu}, \citenamefont {Liu},\ and\
  \citenamefont {Adesso}}]{Takagi19PRL}%
  \BibitemOpen
  \bibfield  {author} {\bibinfo {author} {\bibfnamefont {R.}~\bibnamefont
  {Takagi}}, \bibinfo {author} {\bibfnamefont {B.}~\bibnamefont {Regula}},
  \bibinfo {author} {\bibfnamefont {K.}~\bibnamefont {Bu}}, \bibinfo {author}
  {\bibfnamefont {Z.-W.}\ \bibnamefont {Liu}}, \ and\ \bibinfo {author}
  {\bibfnamefont {G.}~\bibnamefont {Adesso}},\ }\bibfield  {title} {\enquote
  {\bibinfo {title} {Operational advantage of quantum resources in subchannel
  discrimination},}\ }\href {\doibase 10.1103/PhysRevLett.122.140402}
  {\bibfield  {journal} {\bibinfo  {journal} {Phys. Rev. Lett.}\ }\textbf
  {\bibinfo {volume} {122}},\ \bibinfo {pages} {140402} (\bibinfo {year}
  {2019})}\BibitemShut {NoStop}%
\bibitem [{\citenamefont {Takagi}\ and\ \citenamefont
  {Regula}(2019)}]{Takagi19PRX}%
  \BibitemOpen
  \bibfield  {author} {\bibinfo {author} {\bibfnamefont {R.}~\bibnamefont
  {Takagi}}\ and\ \bibinfo {author} {\bibfnamefont {B.}~\bibnamefont
  {Regula}},\ }\bibfield  {title} {\enquote {\bibinfo {title} {General resource
  theories in quantum mechanics and beyond: Operational characterization via
  discrimination tasks},}\ }\href {\doibase 10.1103/PhysRevX.9.031053}
  {\bibfield  {journal} {\bibinfo  {journal} {Phys. Rev. X}\ }\textbf {\bibinfo
  {volume} {9}},\ \bibinfo {pages} {031053} (\bibinfo {year}
  {2019})}\BibitemShut {NoStop}%
\bibitem [{\citenamefont {Oszmaniec}\ and\ \citenamefont
  {Biswas}(2019)}]{Oszmaniec2019operational}%
  \BibitemOpen
  \bibfield  {author} {\bibinfo {author} {\bibfnamefont {M.}~\bibnamefont
  {Oszmaniec}}\ and\ \bibinfo {author} {\bibfnamefont {T.}~\bibnamefont
  {Biswas}},\ }\bibfield  {title} {\enquote {\bibinfo {title} {Operational
  relevance of resource theories of quantum measurements},}\ }\href {\doibase
  10.22331/q-2019-04-26-133} {\bibfield  {journal} {\bibinfo  {journal}
  {{Quantum}}\ }\textbf {\bibinfo {volume} {3}},\ \bibinfo {pages} {133}
  (\bibinfo {year} {2019})}\BibitemShut {NoStop}%
\bibitem [{\citenamefont {Mori}(2020)}]{Mori2019Operational}%
  \BibitemOpen
  \bibfield  {author} {\bibinfo {author} {\bibfnamefont {J.}~\bibnamefont
  {Mori}},\ }\bibfield  {title} {\enquote {\bibinfo {title} {Operational
  characterization of incompatibility of quantum channels with quantum state
  discrimination},}\ }\href {\doibase 10.1103/PhysRevA.101.032331} {\bibfield
  {journal} {\bibinfo  {journal} {Phys. Rev. A}\ }\textbf {\bibinfo {volume}
  {101}},\ \bibinfo {pages} {032331} (\bibinfo {year} {2020})}\BibitemShut
  {NoStop}%
\bibitem [{\citenamefont {Buscemi}\ \emph {et~al.}(2020)\citenamefont
  {Buscemi}, \citenamefont {Chitambar},\ and\ \citenamefont
  {Zhou}}]{Buscemi2020Complete}%
  \BibitemOpen
  \bibfield  {author} {\bibinfo {author} {\bibfnamefont {F.}~\bibnamefont
  {Buscemi}}, \bibinfo {author} {\bibfnamefont {E.}~\bibnamefont {Chitambar}},
  \ and\ \bibinfo {author} {\bibfnamefont {W.}~\bibnamefont {Zhou}},\
  }\bibfield  {title} {\enquote {\bibinfo {title} {Complete resource theory of
  quantum incompatibility as quantum programmability},}\ }\href {\doibase
  10.1103/PhysRevLett.124.120401} {\bibfield  {journal} {\bibinfo  {journal}
  {Phys. Rev. Lett.}\ }\textbf {\bibinfo {volume} {124}},\ \bibinfo {pages}
  {120401} (\bibinfo {year} {2020})}\BibitemShut {NoStop}%
\bibitem [{\citenamefont {Lahti}(2003)}]{Lahti03}%
  \BibitemOpen
  \bibfield  {author} {\bibinfo {author} {\bibfnamefont {P.}~\bibnamefont
  {Lahti}},\ }\href {\doibase 10.1023/a:1025406103210} {\bibfield  {journal}
  {\bibinfo  {journal} {International Journal of Theoretical Physics}\ }\textbf
  {\bibinfo {volume} {42}},\ \bibinfo {pages} {893--906} (\bibinfo {year}
  {2003})}\BibitemShut {NoStop}%
\bibitem [{\citenamefont {Ac\'{\i}n}\ \emph {et~al.}(2007)\citenamefont
  {Ac\'{\i}n}, \citenamefont {Brunner}, \citenamefont {Gisin}, \citenamefont
  {Massar}, \citenamefont {Pironio},\ and\ \citenamefont {Scarani}}]{Acin07}%
  \BibitemOpen
  \bibfield  {author} {\bibinfo {author} {\bibfnamefont {A.}~\bibnamefont
  {Ac\'{\i}n}}, \bibinfo {author} {\bibfnamefont {N.}~\bibnamefont {Brunner}},
  \bibinfo {author} {\bibfnamefont {N.}~\bibnamefont {Gisin}}, \bibinfo
  {author} {\bibfnamefont {S.}~\bibnamefont {Massar}}, \bibinfo {author}
  {\bibfnamefont {S.}~\bibnamefont {Pironio}}, \ and\ \bibinfo {author}
  {\bibfnamefont {V.}~\bibnamefont {Scarani}},\ }\bibfield  {title} {\enquote
  {\bibinfo {title} {Device-independent security of quantum cryptography
  against collective attacks},}\ }\href {\doibase
  10.1103/PhysRevLett.98.230501} {\bibfield  {journal} {\bibinfo  {journal}
  {Phys. Rev. Lett.}\ }\textbf {\bibinfo {volume} {98}},\ \bibinfo {pages}
  {230501} (\bibinfo {year} {2007})}\BibitemShut {NoStop}%
\bibitem [{\citenamefont {Scarani}(2012)}]{Scarani12}%
  \BibitemOpen
  \bibfield  {author} {\bibinfo {author} {\bibfnamefont {V.}~\bibnamefont
  {Scarani}},\ }\bibfield  {title} {\enquote {\bibinfo {title} {The
  device-independent outlook on quantum physics},}\ }\href@noop {} {\bibfield
  {journal} {\bibinfo  {journal} {Acta Phys. Slovaca}\ }\textbf {\bibinfo
  {volume} {62}},\ \bibinfo {pages} {347--409} (\bibinfo {year}
  {2012})}\BibitemShut {NoStop}%
\bibitem [{\citenamefont {Moroder}\ \emph {et~al.}(2013)\citenamefont
  {Moroder}, \citenamefont {Bancal}, \citenamefont {Liang}, \citenamefont
  {Hofmann},\ and\ \citenamefont {G\"uhne}}]{Moroder13}%
  \BibitemOpen
  \bibfield  {author} {\bibinfo {author} {\bibfnamefont {T.}~\bibnamefont
  {Moroder}}, \bibinfo {author} {\bibfnamefont {J.-D.}\ \bibnamefont {Bancal}},
  \bibinfo {author} {\bibfnamefont {Y.-C.}\ \bibnamefont {Liang}}, \bibinfo
  {author} {\bibfnamefont {M.}~\bibnamefont {Hofmann}}, \ and\ \bibinfo
  {author} {\bibfnamefont {O.}~\bibnamefont {G\"uhne}},\ }\bibfield  {title}
  {\enquote {\bibinfo {title} {Device-independent entanglement quantification
  and related applications},}\ }\href {\doibase 10.1103/PhysRevLett.111.030501}
  {\bibfield  {journal} {\bibinfo  {journal} {Phys. Rev. Lett.}\ }\textbf
  {\bibinfo {volume} {111}},\ \bibinfo {pages} {030501} (\bibinfo {year}
  {2013})}\BibitemShut {NoStop}%
\bibitem [{\citenamefont {Pironio}\ \emph
  {et~al.}(2010{\natexlab{a}})\citenamefont {Pironio}, \citenamefont
  {Ac{\'{\i}}n}, \citenamefont {Massar}, \citenamefont {de~la Giroday},
  \citenamefont {Matsukevich}, \citenamefont {Maunz}, \citenamefont
  {Olmschenk}, \citenamefont {Hayes}, \citenamefont {Luo}, \citenamefont
  {Manning},\ and\ \citenamefont {Monroe}}]{Pironio10}%
  \BibitemOpen
  \bibfield  {author} {\bibinfo {author} {\bibfnamefont {S.}~\bibnamefont
  {Pironio}}, \bibinfo {author} {\bibfnamefont {A.}~\bibnamefont
  {Ac{\'{\i}}n}}, \bibinfo {author} {\bibfnamefont {S.}~\bibnamefont {Massar}},
  \bibinfo {author} {\bibfnamefont {A.~B.}\ \bibnamefont {de~la Giroday}},
  \bibinfo {author} {\bibfnamefont {D.~N.}\ \bibnamefont {Matsukevich}},
  \bibinfo {author} {\bibfnamefont {P.}~\bibnamefont {Maunz}}, \bibinfo
  {author} {\bibfnamefont {S.}~\bibnamefont {Olmschenk}}, \bibinfo {author}
  {\bibfnamefont {D.}~\bibnamefont {Hayes}}, \bibinfo {author} {\bibfnamefont
  {L.}~\bibnamefont {Luo}}, \bibinfo {author} {\bibfnamefont {T.~A.}\
  \bibnamefont {Manning}}, \ and\ \bibinfo {author} {\bibfnamefont
  {C.}~\bibnamefont {Monroe}},\ }\bibfield  {title} {\enquote {\bibinfo {title}
  {Random numbers certified by {B}ell's theorem},}\ }\href {\doibase
  10.1038/nature09008} {\bibfield  {journal} {\bibinfo  {journal} {Nature}\
  }\textbf {\bibinfo {volume} {464}},\ \bibinfo {pages} {1021--1024} (\bibinfo
  {year} {2010}{\natexlab{a}})}\BibitemShut {NoStop}%
\bibitem [{\citenamefont {Wiseman}\ \emph {et~al.}(2007)\citenamefont
  {Wiseman}, \citenamefont {Jones},\ and\ \citenamefont {Doherty}}]{Wiseman07}%
  \BibitemOpen
  \bibfield  {author} {\bibinfo {author} {\bibfnamefont {H.~M.}\ \bibnamefont
  {Wiseman}}, \bibinfo {author} {\bibfnamefont {S.~J.}\ \bibnamefont {Jones}},
  \ and\ \bibinfo {author} {\bibfnamefont {A.~C.}\ \bibnamefont {Doherty}},\
  }\bibfield  {title} {\enquote {\bibinfo {title} {Steering, entanglement,
  nonlocality, and the {E}instein-{P}odolsky-{R}osen paradox},}\ }\href
  {\doibase 10.1103/PhysRevLett.98.140402} {\bibfield  {journal} {\bibinfo
  {journal} {Phys. Rev. Lett.}\ }\textbf {\bibinfo {volume} {98}},\ \bibinfo
  {pages} {140402} (\bibinfo {year} {2007})}\BibitemShut {NoStop}%
\bibitem [{\citenamefont {Cavalcanti}\ and\ \citenamefont
  {Skrzypczyk}(2016)}]{Cavalcanti16}%
  \BibitemOpen
  \bibfield  {author} {\bibinfo {author} {\bibfnamefont {D.}~\bibnamefont
  {Cavalcanti}}\ and\ \bibinfo {author} {\bibfnamefont {P.}~\bibnamefont
  {Skrzypczyk}},\ }\bibfield  {title} {\enquote {\bibinfo {title} {Quantitative
  relations between measurement incompatibility, quantum steering, and
  nonlocality},}\ }\href {\doibase 10.1103/PhysRevA.93.052112} {\bibfield
  {journal} {\bibinfo  {journal} {Phys. Rev. A}\ }\textbf {\bibinfo {volume}
  {93}},\ \bibinfo {pages} {052112} (\bibinfo {year} {2016})}\BibitemShut
  {NoStop}%
\bibitem [{\citenamefont {Chen}\ \emph {et~al.}(2016)\citenamefont {Chen},
  \citenamefont {Budroni}, \citenamefont {Liang},\ and\ \citenamefont
  {Chen}}]{CBLC16}%
  \BibitemOpen
  \bibfield  {author} {\bibinfo {author} {\bibfnamefont {S.-L.}\ \bibnamefont
  {Chen}}, \bibinfo {author} {\bibfnamefont {C.}~\bibnamefont {Budroni}},
  \bibinfo {author} {\bibfnamefont {Y.-C.}\ \bibnamefont {Liang}}, \ and\
  \bibinfo {author} {\bibfnamefont {Y.-N.}\ \bibnamefont {Chen}},\ }\bibfield
  {title} {\enquote {\bibinfo {title} {Natural framework for device-independent
  quantification of quantum steerability, measurement incompatibility, and
  self-testing},}\ }\href {\doibase 10.1103/PhysRevLett.116.240401} {\bibfield
  {journal} {\bibinfo  {journal} {Phys. Rev. Lett.}\ }\textbf {\bibinfo
  {volume} {116}},\ \bibinfo {pages} {240401} (\bibinfo {year}
  {2016})}\BibitemShut {NoStop}%
\bibitem [{\citenamefont {Chen}\ \emph {et~al.}(2018)\citenamefont {Chen},
  \citenamefont {Budroni}, \citenamefont {Liang},\ and\ \citenamefont
  {Chen}}]{CBLC18}%
  \BibitemOpen
  \bibfield  {author} {\bibinfo {author} {\bibfnamefont {S.-L.}\ \bibnamefont
  {Chen}}, \bibinfo {author} {\bibfnamefont {C.}~\bibnamefont {Budroni}},
  \bibinfo {author} {\bibfnamefont {Y.-C.}\ \bibnamefont {Liang}}, \ and\
  \bibinfo {author} {\bibfnamefont {Y.-N.}\ \bibnamefont {Chen}},\ }\bibfield
  {title} {\enquote {\bibinfo {title} {Exploring the framework of assemblage
  moment matrices and its applications in device-independent
  characterizations},}\ }\href {\doibase 10.1103/PhysRevA.98.042127} {\bibfield
   {journal} {\bibinfo  {journal} {Phys. Rev. A}\ }\textbf {\bibinfo {volume}
  {98}},\ \bibinfo {pages} {042127} (\bibinfo {year} {2018})}\BibitemShut
  {NoStop}%
\bibitem [{\citenamefont {Gallego}\ \emph {et~al.}(2010)\citenamefont
  {Gallego}, \citenamefont {Brunner}, \citenamefont {Hadley},\ and\
  \citenamefont {Ac\'{\i}n}}]{Gallego10}%
  \BibitemOpen
  \bibfield  {author} {\bibinfo {author} {\bibfnamefont {R.}~\bibnamefont
  {Gallego}}, \bibinfo {author} {\bibfnamefont {N.}~\bibnamefont {Brunner}},
  \bibinfo {author} {\bibfnamefont {C.}~\bibnamefont {Hadley}}, \ and\ \bibinfo
  {author} {\bibfnamefont {A.}~\bibnamefont {Ac\'{\i}n}},\ }\bibfield  {title}
  {\enquote {\bibinfo {title} {Device-independent tests of classical and
  quantum dimensions},}\ }\href {\doibase 10.1103/PhysRevLett.105.230501}
  {\bibfield  {journal} {\bibinfo  {journal} {Phys. Rev. Lett.}\ }\textbf
  {\bibinfo {volume} {105}},\ \bibinfo {pages} {230501} (\bibinfo {year}
  {2010})}\BibitemShut {NoStop}%
\bibitem [{\citenamefont {Hirsch}\ \emph {et~al.}(2018)\citenamefont {Hirsch},
  \citenamefont {Quintino},\ and\ \citenamefont {Brunner}}]{HirschPRA2018}%
  \BibitemOpen
  \bibfield  {author} {\bibinfo {author} {\bibfnamefont {F.}~\bibnamefont
  {Hirsch}}, \bibinfo {author} {\bibfnamefont {M.~T.}\ \bibnamefont
  {Quintino}}, \ and\ \bibinfo {author} {\bibfnamefont {N.}~\bibnamefont
  {Brunner}},\ }\bibfield  {title} {\enquote {\bibinfo {title} {Quantum
  measurement incompatibility does not imply bell nonlocality},}\ }\href
  {\doibase 10.1103/PhysRevA.97.012129} {\bibfield  {journal} {\bibinfo
  {journal} {Phys. Rev. A}\ }\textbf {\bibinfo {volume} {97}},\ \bibinfo
  {pages} {012129} (\bibinfo {year} {2018})}\BibitemShut {NoStop}%
\bibitem [{\citenamefont {Bene}\ and\ \citenamefont
  {V{\'{e}}rtesi}(2018)}]{BeneNJP2018}%
  \BibitemOpen
  \bibfield  {author} {\bibinfo {author} {\bibfnamefont {E.}~\bibnamefont
  {Bene}}\ and\ \bibinfo {author} {\bibfnamefont {T.}~\bibnamefont
  {V{\'{e}}rtesi}},\ }\bibfield  {title} {\enquote {\bibinfo {title}
  {Measurement incompatibility does not give rise to bell violation in
  general},}\ }\href {\doibase 10.1088/1367-2630/aa9ca3} {\bibfield  {journal}
  {\bibinfo  {journal} {New J. Phys.}\ }\textbf {\bibinfo {volume} {20}},\
  \bibinfo {pages} {013021} (\bibinfo {year} {2018})}\BibitemShut {NoStop}%
\bibitem [{\citenamefont {Doherty}\ \emph {et~al.}(2008)\citenamefont
  {Doherty}, \citenamefont {Liang}, \citenamefont {Toner},\ and\ \citenamefont
  {Wehner}}]{Doherty08}%
  \BibitemOpen
  \bibfield  {author} {\bibinfo {author} {\bibfnamefont {A.~C.}\ \bibnamefont
  {Doherty}}, \bibinfo {author} {\bibfnamefont {Y.-C.}\ \bibnamefont {Liang}},
  \bibinfo {author} {\bibfnamefont {B.}~\bibnamefont {Toner}}, \ and\ \bibinfo
  {author} {\bibfnamefont {S.}~\bibnamefont {Wehner}},\ }\bibfield  {title}
  {\enquote {\bibinfo {title} {The quantum moment problem and bounds on
  entangled multi-prover games},}\ }in\ \href {\doibase 10.1109/CCC.2008.26}
  {\emph {\bibinfo {booktitle} {23rd Annu. IEEE Conf. on Comput. Comp, 2008,
  CCC'08}}}\ (\bibinfo {address} {Los Alamitos, CA},\ \bibinfo {year} {2008})\
  pp.\ \bibinfo {pages} {199--210}\BibitemShut {NoStop}%
\bibitem [{\citenamefont {Navascu\'es}\ \emph {et~al.}(2007)\citenamefont
  {Navascu\'es}, \citenamefont {Pironio},\ and\ \citenamefont
  {Ac\'{\i}n}}]{NPA}%
  \BibitemOpen
  \bibfield  {author} {\bibinfo {author} {\bibfnamefont {M.}~\bibnamefont
  {Navascu\'es}}, \bibinfo {author} {\bibfnamefont {S.}~\bibnamefont
  {Pironio}}, \ and\ \bibinfo {author} {\bibfnamefont {A.}~\bibnamefont
  {Ac\'{\i}n}},\ }\bibfield  {title} {\enquote {\bibinfo {title} {Bounding the
  set of quantum correlations},}\ }\href {\doibase
  10.1103/PhysRevLett.98.010401} {\bibfield  {journal} {\bibinfo  {journal}
  {Phys. Rev. Lett.}\ }\textbf {\bibinfo {volume} {98}},\ \bibinfo {pages}
  {010401} (\bibinfo {year} {2007})}\BibitemShut {NoStop}%
\bibitem [{\citenamefont {Pironio}\ \emph
  {et~al.}(2010{\natexlab{b}})\citenamefont {Pironio}, \citenamefont
  {Navascu{\'{e}}s},\ and\ \citenamefont {Ac{\'{\i}}n}}]{Pironio10b}%
  \BibitemOpen
  \bibfield  {author} {\bibinfo {author} {\bibfnamefont {S.}~\bibnamefont
  {Pironio}}, \bibinfo {author} {\bibfnamefont {M.}~\bibnamefont
  {Navascu{\'{e}}s}}, \ and\ \bibinfo {author} {\bibfnamefont {A.}~\bibnamefont
  {Ac{\'{\i}}n}},\ }\bibfield  {title} {\enquote {\bibinfo {title} {Convergent
  relaxations of polynomial optimization problems with noncommuting
  variables},}\ }\href {\doibase 10.1137/090760155} {\bibfield  {journal}
  {\bibinfo  {journal} {{SIAM} Journal on Optimization}\ }\textbf {\bibinfo
  {volume} {20}},\ \bibinfo {pages} {2157--2180} (\bibinfo {year}
  {2010}{\natexlab{b}})}\BibitemShut {NoStop}%
\bibitem [{\citenamefont {Boyd}\ and\ \citenamefont
  {Vandenberghe}(2004)}]{BoydBook}%
  \BibitemOpen
  \bibfield  {author} {\bibinfo {author} {\bibfnamefont {S.}~\bibnamefont
  {Boyd}}\ and\ \bibinfo {author} {\bibfnamefont {L.}~\bibnamefont
  {Vandenberghe}},\ }\href@noop {} {\emph {\bibinfo {title} {Convex
  Optimization}}},\ \bibinfo {edition} {1st}\ ed.\ (\bibinfo  {publisher}
  {Cambridge University Press, Cambridge},\ \bibinfo {year} {2004})\BibitemShut
  {NoStop}%
\bibitem [{\citenamefont {Haapasalo}(2015)}]{Haapasalo15Robustness}%
  \BibitemOpen
  \bibfield  {author} {\bibinfo {author} {\bibfnamefont {E.}~\bibnamefont
  {Haapasalo}},\ }\bibfield  {title} {\enquote {\bibinfo {title} {Robustness of
  incompatibility for quantum devices},}\ }\href {\doibase
  10.1088/1751-8113/48/25/255303} {\bibfield  {journal} {\bibinfo  {journal}
  {Journal of Physics A: Mathematical and Theoretical}\ }\textbf {\bibinfo
  {volume} {48}},\ \bibinfo {pages} {255303} (\bibinfo {year}
  {2015})}\BibitemShut {NoStop}%
\bibitem [{\citenamefont {Uola}\ \emph {et~al.}(2015)\citenamefont {Uola},
  \citenamefont {Budroni}, \citenamefont {G\"uhne},\ and\ \citenamefont
  {Pellonp\"a\"a}}]{Uola15}%
  \BibitemOpen
  \bibfield  {author} {\bibinfo {author} {\bibfnamefont {R.}~\bibnamefont
  {Uola}}, \bibinfo {author} {\bibfnamefont {C.}~\bibnamefont {Budroni}},
  \bibinfo {author} {\bibfnamefont {O.}~\bibnamefont {G\"uhne}}, \ and\
  \bibinfo {author} {\bibfnamefont {J.-P.}\ \bibnamefont {Pellonp\"a\"a}},\
  }\bibfield  {title} {\enquote {\bibinfo {title} {One-to-one mapping between
  steering and joint measurability problems},}\ }\href {\doibase
  10.1103/PhysRevLett.115.230402} {\bibfield  {journal} {\bibinfo  {journal}
  {Phys. Rev. Lett.}\ }\textbf {\bibinfo {volume} {115}},\ \bibinfo {pages}
  {230402} (\bibinfo {year} {2015})}\BibitemShut {NoStop}%
\bibitem [{\citenamefont {Heinosaari}\ \emph {et~al.}(2016)\citenamefont
  {Heinosaari}, \citenamefont {Miyadera},\ and\ \citenamefont
  {Ziman}}]{Heinosaari16}%
  \BibitemOpen
  \bibfield  {author} {\bibinfo {author} {\bibfnamefont {T.}~\bibnamefont
  {Heinosaari}}, \bibinfo {author} {\bibfnamefont {T.}~\bibnamefont
  {Miyadera}}, \ and\ \bibinfo {author} {\bibfnamefont {M.}~\bibnamefont
  {Ziman}},\ }\bibfield  {title} {\enquote {\bibinfo {title} {An invitation to
  quantum incompatibility},}\ }\href {\doibase 10.1088/1751-8113/49/12/123001}
  {\bibfield  {journal} {\bibinfo  {journal} {Journal of Physics A:
  Mathematical and Theoretical}\ }\textbf {\bibinfo {volume} {49}},\ \bibinfo
  {pages} {123001} (\bibinfo {year} {2016})}\BibitemShut {NoStop}%
\bibitem [{\citenamefont {Quintino}\ \emph {et~al.}(2019)\citenamefont
  {Quintino}, \citenamefont {Budroni}, \citenamefont {Woodhead}, \citenamefont
  {Cabello},\ and\ \citenamefont {Cavalcanti}}]{QuintinoPRL2019}%
  \BibitemOpen
  \bibfield  {author} {\bibinfo {author} {\bibfnamefont {M.~T.}\ \bibnamefont
  {Quintino}}, \bibinfo {author} {\bibfnamefont {C.}~\bibnamefont {Budroni}},
  \bibinfo {author} {\bibfnamefont {E.}~\bibnamefont {Woodhead}}, \bibinfo
  {author} {\bibfnamefont {A.}~\bibnamefont {Cabello}}, \ and\ \bibinfo
  {author} {\bibfnamefont {D.}~\bibnamefont {Cavalcanti}},\ }\bibfield  {title}
  {\enquote {\bibinfo {title} {Device-independent tests of structures of
  measurement incompatibility},}\ }\href {\doibase
  10.1103/PhysRevLett.123.180401} {\bibfield  {journal} {\bibinfo  {journal}
  {Phys. Rev. Lett.}\ }\textbf {\bibinfo {volume} {123}},\ \bibinfo {pages}
  {180401} (\bibinfo {year} {2019})}\BibitemShut {NoStop}%
\bibitem [{\citenamefont {Pusey}(2015)}]{Pusey15}%
  \BibitemOpen
  \bibfield  {author} {\bibinfo {author} {\bibfnamefont {M.~F.}\ \bibnamefont
  {Pusey}},\ }\bibfield  {title} {\enquote {\bibinfo {title} {Verifying the
  quantumness of a channel with an untrusted device},}\ }\href {\doibase
  10.1364/JOSAB.32.000A56} {\bibfield  {journal} {\bibinfo  {journal} {J. Opt.
  Soc. Am. B}\ }\textbf {\bibinfo {volume} {32}},\ \bibinfo {pages} {A56--A63}
  (\bibinfo {year} {2015})}\BibitemShut {NoStop}%
\bibitem [{\citenamefont {Heinosaari}\ \emph {et~al.}(2015)\citenamefont
  {Heinosaari}, \citenamefont {Kiukas},\ and\ \citenamefont
  {Reitzner}}]{Heinosaari15}%
  \BibitemOpen
  \bibfield  {author} {\bibinfo {author} {\bibfnamefont {T.}~\bibnamefont
  {Heinosaari}}, \bibinfo {author} {\bibfnamefont {J.}~\bibnamefont {Kiukas}},
  \ and\ \bibinfo {author} {\bibfnamefont {D.}~\bibnamefont {Reitzner}},\
  }\bibfield  {title} {\enquote {\bibinfo {title} {Noise robustness of the
  incompatibility of quantum measurements},}\ }\href {\doibase
  10.1103/PhysRevA.92.022115} {\bibfield  {journal} {\bibinfo  {journal} {Phys.
  Rev. A}\ }\textbf {\bibinfo {volume} {92}},\ \bibinfo {pages} {022115}
  (\bibinfo {year} {2015})}\BibitemShut {NoStop}%
\bibitem [{\citenamefont {Paw\l{}owski}\ and\ \citenamefont
  {Brunner}(2011)}]{PawlowskiPRA2011}%
  \BibitemOpen
  \bibfield  {author} {\bibinfo {author} {\bibfnamefont {M.}~\bibnamefont
  {Paw\l{}owski}}\ and\ \bibinfo {author} {\bibfnamefont {N.}~\bibnamefont
  {Brunner}},\ }\bibfield  {title} {\enquote {\bibinfo {title}
  {Semi-device-independent security of one-way quantum key distribution},}\
  }\href {\doibase 10.1103/PhysRevA.84.010302} {\bibfield  {journal} {\bibinfo
  {journal} {Phys. Rev. A}\ }\textbf {\bibinfo {volume} {84}},\ \bibinfo
  {pages} {010302} (\bibinfo {year} {2011})}\BibitemShut {NoStop}%
\bibitem [{\citenamefont {Liang}\ \emph
  {et~al.}(2011{\natexlab{b}})\citenamefont {Liang}, \citenamefont
  {V\'ertesi},\ and\ \citenamefont {Brunner}}]{LiangPRA2011}%
  \BibitemOpen
  \bibfield  {author} {\bibinfo {author} {\bibfnamefont {Y.-C.}\ \bibnamefont
  {Liang}}, \bibinfo {author} {\bibfnamefont {T.}~\bibnamefont {V\'ertesi}}, \
  and\ \bibinfo {author} {\bibfnamefont {N.}~\bibnamefont {Brunner}},\
  }\bibfield  {title} {\enquote {\bibinfo {title} {Semi-device-independent
  bounds on entanglement},}\ }\href {\doibase 10.1103/PhysRevA.83.022108}
  {\bibfield  {journal} {\bibinfo  {journal} {Phys. Rev. A}\ }\textbf {\bibinfo
  {volume} {83}},\ \bibinfo {pages} {022108} (\bibinfo {year}
  {2011}{\natexlab{b}})}\BibitemShut {NoStop}%
\bibitem [{\citenamefont {Piani}\ and\ \citenamefont
  {Watrous}(2015)}]{Piani15}%
  \BibitemOpen
  \bibfield  {author} {\bibinfo {author} {\bibfnamefont {M.}~\bibnamefont
  {Piani}}\ and\ \bibinfo {author} {\bibfnamefont {J.}~\bibnamefont
  {Watrous}},\ }\bibfield  {title} {\enquote {\bibinfo {title} {Necessary and
  sufficient quantum information characterization of
  {E}instein-{P}odolsky-{R}osen steering},}\ }\href {\doibase
  10.1103/PhysRevLett.114.060404} {\bibfield  {journal} {\bibinfo  {journal}
  {Phys. Rev. Lett.}\ }\textbf {\bibinfo {volume} {114}},\ \bibinfo {pages}
  {060404} (\bibinfo {year} {2015})}\BibitemShut {NoStop}%
\bibitem [{\citenamefont {Busch}\ \emph {et~al.}(1996)\citenamefont {Busch},
  \citenamefont {Lahti},\ and\ \citenamefont {Mittelstaedt}}]{BuschBook}%
  \BibitemOpen
  \bibfield  {author} {\bibinfo {author} {\bibfnamefont {P.}~\bibnamefont
  {Busch}}, \bibinfo {author} {\bibfnamefont {P.~J.}\ \bibnamefont {Lahti}}, \
  and\ \bibinfo {author} {\bibfnamefont {P.}~\bibnamefont {Mittelstaedt}},\
  }\href@noop {} {\emph {\bibinfo {title} {The Quantum Theory of
  Measurement}}},\ \bibinfo {edition} {2nd}\ ed.,\ \bibinfo {series} {Lecture
  Notes in Physics Monographs}, Vol.~\bibinfo {volume} {2}\ (\bibinfo
  {publisher} {Springer-Verlag Berlin Heidelberg},\ \bibinfo {year}
  {1996})\BibitemShut {NoStop}%
\bibitem [{\citenamefont {Ali}\ \emph {et~al.}(2009)\citenamefont {Ali},
  \citenamefont {Carmeli}, \citenamefont {Heinosaari},\ and\ \citenamefont
  {Toigo}}]{Ali09}%
  \BibitemOpen
  \bibfield  {author} {\bibinfo {author} {\bibfnamefont {S.~T.}\ \bibnamefont
  {Ali}}, \bibinfo {author} {\bibfnamefont {C.}~\bibnamefont {Carmeli}},
  \bibinfo {author} {\bibfnamefont {T.}~\bibnamefont {Heinosaari}}, \ and\
  \bibinfo {author} {\bibfnamefont {A.}~\bibnamefont {Toigo}},\ }\bibfield
  {title} {\enquote {\bibinfo {title} {Commutative {POVMs} and fuzzy
  observables},}\ }\href {\doibase 10.1007/s10701-009-9292-y} {\bibfield
  {journal} {\bibinfo  {journal} {Foundations of Physics}\ }\textbf {\bibinfo
  {volume} {39}},\ \bibinfo {pages} {593--612} (\bibinfo {year}
  {2009})}\BibitemShut {NoStop}%
\bibitem [{\citenamefont {Designolle}\ \emph {et~al.}(2019)\citenamefont
  {Designolle}, \citenamefont {Farkas},\ and\ \citenamefont
  {Kaniewski}}]{Designolle19}%
  \BibitemOpen
  \bibfield  {author} {\bibinfo {author} {\bibfnamefont {S.}~\bibnamefont
  {Designolle}}, \bibinfo {author} {\bibfnamefont {M.}~\bibnamefont {Farkas}},
  \ and\ \bibinfo {author} {\bibfnamefont {J.}~\bibnamefont {Kaniewski}},\
  }\bibfield  {title} {\enquote {\bibinfo {title} {Incompatibility robustness
  of quantum measurements: a unified framework},}\ }\href {\doibase
  10.1088/1367-2630/ab5020} {\bibfield  {journal} {\bibinfo  {journal} {New
  Journal of Physics}\ }\textbf {\bibinfo {volume} {21}},\ \bibinfo {pages}
  {113053} (\bibinfo {year} {2019})}\BibitemShut {NoStop}%
\bibitem [{\citenamefont {Navascu{\'e}s}\ \emph {et~al.}(2008)\citenamefont
  {Navascu{\'e}s}, \citenamefont {Pironio},\ and\ \citenamefont
  {Ac{\'\i}n}}]{NPA2008}%
  \BibitemOpen
  \bibfield  {author} {\bibinfo {author} {\bibfnamefont {M.}~\bibnamefont
  {Navascu{\'e}s}}, \bibinfo {author} {\bibfnamefont {S.}~\bibnamefont
  {Pironio}}, \ and\ \bibinfo {author} {\bibfnamefont {A.}~\bibnamefont
  {Ac{\'\i}n}},\ }\bibfield  {title} {\enquote {\bibinfo {title} {A convergent
  hierarchy of semidefinite programs characterizing the set of quantum
  correlations},}\ }\href {http://stacks.iop.org/1367-2630/10/i=7/a=073013}
  {\bibfield  {journal} {\bibinfo  {journal} {New Journal of Physics}\ }\textbf
  {\bibinfo {volume} {10}},\ \bibinfo {pages} {073013} (\bibinfo {year}
  {2008})}\BibitemShut {NoStop}%
\bibitem [{\citenamefont {Peres}(1990)}]{Peres90}%
  \BibitemOpen
  \bibfield  {author} {\bibinfo {author} {\bibfnamefont {A.}~\bibnamefont
  {Peres}},\ }\bibfield  {title} {\enquote {\bibinfo {title} {Neumark's theorem
  and quantum inseparability},}\ }\href {\doibase 10.1007/bf01883517}
  {\bibfield  {journal} {\bibinfo  {journal} {Foundations of Physics}\ }\textbf
  {\bibinfo {volume} {20}},\ \bibinfo {pages} {1441--1453} (\bibinfo {year}
  {1990})}\BibitemShut {NoStop}%
\bibitem [{\citenamefont {Ac\'{\i}n}\ \emph {et~al.}(2012)\citenamefont
  {Ac\'{\i}n}, \citenamefont {Massar},\ and\ \citenamefont {Pironio}}]{Acin12}%
  \BibitemOpen
  \bibfield  {author} {\bibinfo {author} {\bibfnamefont {A.}~\bibnamefont
  {Ac\'{\i}n}}, \bibinfo {author} {\bibfnamefont {S.}~\bibnamefont {Massar}}, \
  and\ \bibinfo {author} {\bibfnamefont {S.}~\bibnamefont {Pironio}},\
  }\bibfield  {title} {\enquote {\bibinfo {title} {Randomness versus
  nonlocality and entanglement},}\ }\href {\doibase
  10.1103/PhysRevLett.108.100402} {\bibfield  {journal} {\bibinfo  {journal}
  {Phys. Rev. Lett.}\ }\textbf {\bibinfo {volume} {108}},\ \bibinfo {pages}
  {100402} (\bibinfo {year} {2012})}\BibitemShut {NoStop}%
\bibitem [{\citenamefont {Yang}\ and\ \citenamefont
  {Navascu\'es}(2013)}]{Yang13}%
  \BibitemOpen
  \bibfield  {author} {\bibinfo {author} {\bibfnamefont {T.~H.}\ \bibnamefont
  {Yang}}\ and\ \bibinfo {author} {\bibfnamefont {M.}~\bibnamefont
  {Navascu\'es}},\ }\bibfield  {title} {\enquote {\bibinfo {title} {Robust
  self-testing of unknown quantum systems into any entangled two-qubit
  states},}\ }\href {\doibase 10.1103/PhysRevA.87.050102} {\bibfield  {journal}
  {\bibinfo  {journal} {Phys. Rev. A}\ }\textbf {\bibinfo {volume} {87}},\
  \bibinfo {pages} {050102} (\bibinfo {year} {2013})}\BibitemShut {NoStop}%
\bibitem [{\citenamefont {Bamps}\ and\ \citenamefont
  {Pironio}(2015)}]{Bamps15}%
  \BibitemOpen
  \bibfield  {author} {\bibinfo {author} {\bibfnamefont {C.}~\bibnamefont
  {Bamps}}\ and\ \bibinfo {author} {\bibfnamefont {S.}~\bibnamefont
  {Pironio}},\ }\bibfield  {title} {\enquote {\bibinfo {title} {Sum-of-squares
  decompositions for a family of {C}lauser-{H}orne-{S}himony-{H}olt-like
  inequalities and their application to self-testing},}\ }\href {\doibase
  10.1103/PhysRevA.91.052111} {\bibfield  {journal} {\bibinfo  {journal} {Phys.
  Rev. A}\ }\textbf {\bibinfo {volume} {91}},\ \bibinfo {pages} {052111}
  (\bibinfo {year} {2015})}\BibitemShut {NoStop}%
\bibitem [{\citenamefont {Collins}\ and\ \citenamefont
  {Gisin}(2004)}]{Collins04}%
  \BibitemOpen
  \bibfield  {author} {\bibinfo {author} {\bibfnamefont {D.}~\bibnamefont
  {Collins}}\ and\ \bibinfo {author} {\bibfnamefont {N.}~\bibnamefont
  {Gisin}},\ }\bibfield  {title} {\enquote {\bibinfo {title} {A relevant two
  qubit {B}ell inequality inequivalent to the {CHSH} inequality},}\ }\href
  {http://stacks.iop.org/0305-4470/37/i=5/a=021} {\bibfield  {journal}
  {\bibinfo  {journal} {J. Phys. A: Math. Theo.}\ }\textbf {\bibinfo {volume}
  {37}},\ \bibinfo {pages} {1775} (\bibinfo {year} {2004})}\BibitemShut
  {NoStop}%
\bibitem [{\citenamefont {Navascu\'es}\ and\ \citenamefont
  {V\'ertesi}(2015)}]{Navascues15prl}%
  \BibitemOpen
  \bibfield  {author} {\bibinfo {author} {\bibfnamefont {M.}~\bibnamefont
  {Navascu\'es}}\ and\ \bibinfo {author} {\bibfnamefont {T.}~\bibnamefont
  {V\'ertesi}},\ }\bibfield  {title} {\enquote {\bibinfo {title} {Bounding the
  set of finite dimensional quantum correlations},}\ }\href {\doibase
  10.1103/PhysRevLett.115.020501} {\bibfield  {journal} {\bibinfo  {journal}
  {Phys. Rev. Lett.}\ }\textbf {\bibinfo {volume} {115}},\ \bibinfo {pages}
  {020501} (\bibinfo {year} {2015})}\BibitemShut {NoStop}%
\bibitem [{\citenamefont {Kaniewski}\ \emph {et~al.}(2019)\citenamefont
  {Kaniewski}, \citenamefont {{\v{S}}upi{\'{c}}}, \citenamefont {Tura},
  \citenamefont {Baccari}, \citenamefont {Salavrakos},\ and\ \citenamefont
  {Augusiak}}]{Kaniewski19}%
  \BibitemOpen
  \bibfield  {author} {\bibinfo {author} {\bibfnamefont {J.}~\bibnamefont
  {Kaniewski}}, \bibinfo {author} {\bibfnamefont {I.}~\bibnamefont
  {{\v{S}}upi{\'{c}}}}, \bibinfo {author} {\bibfnamefont {J.}~\bibnamefont
  {Tura}}, \bibinfo {author} {\bibfnamefont {F.}~\bibnamefont {Baccari}},
  \bibinfo {author} {\bibfnamefont {A.}~\bibnamefont {Salavrakos}}, \ and\
  \bibinfo {author} {\bibfnamefont {R.}~\bibnamefont {Augusiak}},\ }\bibfield
  {title} {\enquote {\bibinfo {title} {Maximal nonlocality from maximal
  entanglement and mutually unbiased bases, and self-testing of two-qutrit
  quantum systems},}\ }\href {\doibase 10.22331/q-2019-10-24-198} {\bibfield
  {journal} {\bibinfo  {journal} {{Quantum}}\ }\textbf {\bibinfo {volume}
  {3}},\ \bibinfo {pages} {198} (\bibinfo {year} {2019})}\BibitemShut {NoStop}%
\bibitem [{\citenamefont {Tavakoli}\ \emph {et~al.}(2019)\citenamefont
  {Tavakoli}, \citenamefont {Farkas}, \citenamefont {Rosset}, \citenamefont
  {Bancal},\ and\ \citenamefont {Kaniewski}}]{Tavakoli19}%
  \BibitemOpen
  \bibfield  {author} {\bibinfo {author} {\bibfnamefont {A.}~\bibnamefont
  {Tavakoli}}, \bibinfo {author} {\bibfnamefont {M.}~\bibnamefont {Farkas}},
  \bibinfo {author} {\bibfnamefont {D.}~\bibnamefont {Rosset}}, \bibinfo
  {author} {\bibfnamefont {J.-D.}\ \bibnamefont {Bancal}}, \ and\ \bibinfo
  {author} {\bibfnamefont {J.}~\bibnamefont {Kaniewski}},\ }\href@noop {}
  {\enquote {\bibinfo {title} {Mutually unbiased bases and symmetric
  informationally complete measurements in bell experiments: Bell inequalities,
  device-independent certification and applications},}\ } (\bibinfo {year}
  {2019}),\ \Eprint {http://arxiv.org/abs/arXiv:1912.03225} {arXiv:1912.03225}
  \BibitemShut {NoStop}%
\bibitem [{\citenamefont {Chen}\ \emph {et~al.}(2020)\citenamefont {Chen},
  \citenamefont {Ku}, \citenamefont {Zhou}, \citenamefont {Tura},\ and\
  \citenamefont {Chen}}]{Chen2020Robust}%
  \BibitemOpen
  \bibfield  {author} {\bibinfo {author} {\bibfnamefont {S.-L.}\ \bibnamefont
  {Chen}}, \bibinfo {author} {\bibfnamefont {H.-Y.}\ \bibnamefont {Ku}},
  \bibinfo {author} {\bibfnamefont {W.}~\bibnamefont {Zhou}}, \bibinfo {author}
  {\bibfnamefont {J.}~\bibnamefont {Tura}}, \ and\ \bibinfo {author}
  {\bibfnamefont {Y.-N.}\ \bibnamefont {Chen}},\ }\href@noop {} {\enquote
  {\bibinfo {title} {Robust self-testing of steerable quantum assemblages and
  its applications on device-independent quantum certification},}\ } (\bibinfo
  {year} {2020}),\ \Eprint {http://arxiv.org/abs/arXiv:2002.02823}
  {arXiv:2002.02823} \BibitemShut {NoStop}%
\bibitem [{\citenamefont {Carmeli}\ \emph {et~al.}(2020)\citenamefont
  {Carmeli}, \citenamefont {Heinosaari},\ and\ \citenamefont
  {Toigo}}]{Carmeli_2020}%
  \BibitemOpen
  \bibfield  {author} {\bibinfo {author} {\bibfnamefont {C.}~\bibnamefont
  {Carmeli}}, \bibinfo {author} {\bibfnamefont {T.}~\bibnamefont {Heinosaari}},
  \ and\ \bibinfo {author} {\bibfnamefont {A.}~\bibnamefont {Toigo}},\
  }\bibfield  {title} {\enquote {\bibinfo {title} {Quantum random access codes
  and incompatibility of measurements},}\ }\href {\doibase
  10.1209/0295-5075/130/50001} {\bibfield  {journal} {\bibinfo  {journal}
  {{EPL} (Europhysics Letters)}\ }\textbf {\bibinfo {volume} {130}},\ \bibinfo
  {pages} {50001} (\bibinfo {year} {2020})}\BibitemShut {NoStop}%
\bibitem [{\citenamefont {Heinosaari}\ \emph {et~al.}(2014)\citenamefont
  {Heinosaari}, \citenamefont {Schultz}, \citenamefont {Toigo},\ and\
  \citenamefont {Ziman}}]{Heinosaari14}%
  \BibitemOpen
  \bibfield  {author} {\bibinfo {author} {\bibfnamefont {T.}~\bibnamefont
  {Heinosaari}}, \bibinfo {author} {\bibfnamefont {J.}~\bibnamefont {Schultz}},
  \bibinfo {author} {\bibfnamefont {A.}~\bibnamefont {Toigo}}, \ and\ \bibinfo
  {author} {\bibfnamefont {M.}~\bibnamefont {Ziman}},\ }\bibfield  {title}
  {\enquote {\bibinfo {title} {Maximally incompatible quantum observables},}\
  }\href {\doibase 10.1016/j.physleta.2014.04.026} {\bibfield  {journal}
  {\bibinfo  {journal} {Physics Letters A}\ }\textbf {\bibinfo {volume}
  {378}},\ \bibinfo {pages} {1695--1699} (\bibinfo {year} {2014})}\BibitemShut
  {NoStop}%
\bibitem [{\citenamefont {Pironio}(2014)}]{Pironio_2014}%
  \BibitemOpen
  \bibfield  {author} {\bibinfo {author} {\bibfnamefont {S.}~\bibnamefont
  {Pironio}},\ }\bibfield  {title} {\enquote {\bibinfo {title} {All
  clauser{\textendash}horne{\textendash}shimony{\textendash}holt polytopes},}\
  }\href {\doibase 10.1088/1751-8113/47/42/424020} {\bibfield  {journal}
  {\bibinfo  {journal} {Journal of Physics A: Mathematical and Theoretical}\
  }\textbf {\bibinfo {volume} {47}},\ \bibinfo {pages} {424020} (\bibinfo
  {year} {2014})}\BibitemShut {NoStop}%
\bibitem [{\citenamefont {Rosset}(2018)}]{Rosset18SymDPoly}%
  \BibitemOpen
  \bibfield  {author} {\bibinfo {author} {\bibfnamefont {D.}~\bibnamefont
  {Rosset}},\ }\href@noop {} {\enquote {\bibinfo {title} {Symdpoly:
  symmetry-adapted moment relaxations for noncommutative polynomial
  optimization},}\ } (\bibinfo {year} {2018}),\ \Eprint
  {http://arxiv.org/abs/arXiv:1808.09598} {arXiv:1808.09598} \BibitemShut
  {NoStop}%
\bibitem [{\citenamefont {Hoffmann}\ \emph {et~al.}(2018)\citenamefont
  {Hoffmann}, \citenamefont {Spee}, \citenamefont {Gühne},\ and\ \citenamefont
  {Budroni}}]{Hoffmann_2018}%
  \BibitemOpen
  \bibfield  {author} {\bibinfo {author} {\bibfnamefont {J.}~\bibnamefont
  {Hoffmann}}, \bibinfo {author} {\bibfnamefont {C.}~\bibnamefont {Spee}},
  \bibinfo {author} {\bibfnamefont {O.}~\bibnamefont {Gühne}}, \ and\ \bibinfo
  {author} {\bibfnamefont {C.}~\bibnamefont {Budroni}},\ }\bibfield  {title}
  {\enquote {\bibinfo {title} {Structure of temporal correlations of a
  qubit},}\ }\href {\doibase 10.1088/1367-2630/aae87f} {\bibfield  {journal}
  {\bibinfo  {journal} {New Journal of Physics}\ }\textbf {\bibinfo {volume}
  {20}},\ \bibinfo {pages} {102001} (\bibinfo {year} {2018})}\BibitemShut
  {NoStop}%
\bibitem [{\citenamefont {Budroni}\ \emph {et~al.}(2019)\citenamefont
  {Budroni}, \citenamefont {Fagundes},\ and\ \citenamefont
  {Kleinmann}}]{Budroni_2019}%
  \BibitemOpen
  \bibfield  {author} {\bibinfo {author} {\bibfnamefont {C.}~\bibnamefont
  {Budroni}}, \bibinfo {author} {\bibfnamefont {G.}~\bibnamefont {Fagundes}}, \
  and\ \bibinfo {author} {\bibfnamefont {M.}~\bibnamefont {Kleinmann}},\
  }\bibfield  {title} {\enquote {\bibinfo {title} {Memory cost of temporal
  correlations},}\ }\href {\doibase 10.1088/1367-2630/ab3cb4} {\bibfield
  {journal} {\bibinfo  {journal} {New Journal of Physics}\ }\textbf {\bibinfo
  {volume} {21}},\ \bibinfo {pages} {093018} (\bibinfo {year}
  {2019})}\BibitemShut {NoStop}%
\end{thebibliography}

\end{document}